# Fluxional Behavior at the Atomic Level and its Impact on Activity: CO Oxidation over CeO$_2$-supported Pt Catalysts

Joshua L. Vincent[1], and Peter A. Crozier[1]*

[1] *School for Engineering of Matter, Transport, and Energy, Arizona State University, Tempe, Arizona 85281*

*Corresponding author email: crozier@asu.edu

## Abstract

Reducible oxides are widely used catalyst supports that can increase oxidation reaction rates by transferring their lattice oxygen at the metal-support interface. The interfacial oxidation process is typically described in terms of a Mars-van Krevelen mechanism. However, many outstanding questions remain unanswered regarding the atomic-scale structure and dynamic meta-stability (i.e., fluxional behavior) of the interface *during catalysis*. Here, we employ *operando* transmission electron microscopy (TEM) to visualize the structural dynamics occurring at and near $Pt/CeO_2$ interfaces during CO oxidation. Finite element modeling is performed to develop a reaction rate analysis wherein the atomic-level structural observations are directly correlated with the catalyst's turnover frequency for CO oxidation. We show that the increasing frequency of catalytic turnover correlates with dynamic fluxional behavior that (a) destabilizes the supported Pt particle, (b) marks an enhanced rate of oxygen vacancy creation and annihilation, and (c) leads to increased strain and reduction in the surface of the $CeO_2$ support. Overall, the results implicate the interfacial Pt-O-Ce bonds anchoring the Pt to the support as being involved also in the catalytically-driven oxygen transfer process, and they suggest that oxygen reduction takes place on the highly reduced nearby $CeO_2$ surface before migrating to the interfacial perimeter for reaction with CO. The *operando* electron microscopy approach described here should be applicable to a large number of nanoparticle catalysts. This technique will enable the identification of catalytically functional surface structures and strengthen our ability to establish (dynamic) structure-activity relations.

# 1. Introduction

Reducible oxides such as $CeO_2$ are widely used catalyst supports due to their ability to undergo rapid and reversible oxygen uptake and release (Aneggi et al., 2016; Trovarelli, 2002; Li et al., 2019). Additionally, reducible oxide supports offer strong metal-support interactions and can enhance the rates of oxidation reactions by transferring their lattice oxygen to reactive adsorbates at the metal-support interface (Datye et al., 1995; Carrettin et al., 2004; Cargnello et al., 2013; Kopelent et al., 2015; Pereira-Hernández et al., 2019; Lu et al., 2020). For CO oxidation, the interfacial oxidation process is typically described in terms of a Mars-van Krevelen mechanism (Mars & van Krevelen, 1954; Puigdollers et al., 2017), in which $CeO_2$ lattice oxygen is transferred to CO at the perimeter of the metal-support interface, which is also called the three-phase boundary. There is very little information about the atomic structure and structural dynamics of an active metal-support interface performing catalysis. Consequently, there are many outstanding questions regarding the atomic scale evolution of the interface, metal particle, and adjacent oxide surface during catalysis. For example, how does the metal particle and the metal-support interface behave during catalysis if oxygen vacancies are constantly being created and annihilated? Does such a process destabilize the metal particles and make coarsening more likely? Where is the likely site for molecular oxygen reduction and reincorporation into the lattice, which completes the catalytic cycle?

One may expect substantial changes in metal particle shape and bonding to occur during catalytic turnover, as the interfacial energy fluctuates from the continual creation and annihilation of oxygen vacancies. The adhesion between the metal particle and support may also weaken significantly since bridging interfacial oxygen are responsible for anchoring the metal to the support (Hatanaka et al., 2010; Shinjoh et al., 2009; Nagai et al., 2006; Gänzler et al., 2017, 2018).

Oxygen vacancies will introduce strain in the $CeO_2$, which could further destabilize the system and make it more reactive (Balaji Gopal et al., 2017). Recently, there has been an emerging paradigm that has roots in surface science (Somorjai, 1991) and chemistry (Cotton, 1975) for understanding catalytically active sites as dynamic, meta-stable, or so-called fluxional species (Zhang et al., 2020; Sun & Sautet, 2018; Zhai & Alexandrova, 2017; Ethan L Lawrence et al., 2021; Li et al., 2021). In order to deepen our understanding of the factors affecting catalysis and to develop strategies for improved catalyst design, it is essential to elucidate and describe the structural meta-stability (i.e., fluxional behavior) that occurs at the atomic level during simultaneous catalytic turnover.

Aberration-corrected *in situ* environmental transmission electron microscopy (AC-ETEM) is a powerful tool capable of providing atomic-level information about dynamic structures and how they evolve during catalysis. In the last decade there have been considerable advancements for studies under reaction conditions, including the development of in-line gas analysis (e.g., spectroscopy and/or mass spectrometry). Measurements of the gas composition can confirm that catalysis is actually taking place, and in favorable cases they may be used to quantify the *in situ* conversion. Recent investigations have used AC-ETEM along with *in situ* microreactors coupled to in-line spectroscopy or downstream mass spectrometry in order to correlate observations of catalyst structure with changes in gas composition or conversion (Vendelbo et al., 2014; Li et al., 2015; Chee et al., 2020; Plodinec et al., 2020; Miller & Crozier, 2021). While impressive, these and other investigations have often involved the observation of larger 10 – 100 nm metal nanoparticles that are dispersed on non-reducible supports (e.g., $SiO_2$) or on $SiN_x$ films. In order to understand Mars-van Krevelen oxidation, however, the metal nanoparticles must be supported on a reducible oxide, since strong metal-support interactions play a significant role in the reaction mechanism. To our knowledge, no such observations have been reported.

Additionally, in order to establish a truly *operando* connection between the observed structure and the catalyst's activity, it is necessary to move beyond simple *in situ* measurements of reactant conversion and directly correlate the observations with the *reaction rate* of the working catalyst, as first discussed by Bañares et al. (Guerrero-Pérez & Bañares, 2006; Bañares, 2005). It is important to stress that an *in situ* measurement of conversion is not itself and in most cases is not directly related to the reaction rate of the working catalyst. In the field of reaction engineering, this knowledge has led to the design of controlled catalytic reactors operated under easily modeled conditions wherein the conversion can be directly related to the kinetics of the catalyzed chemical reaction. The *in situ* reactors used for ETEM catalysis research are complex but may nonetheless be modeled using, e.g., finite element approaches, enabling the determination of quantitative catalytic reaction rates during atomic-level structural observation (Vincent et al., 2020).

In this work, we implement a combined experimental and finite element modeling approach to investigate the atomic-level fluxional dynamics that occur at and near $Pt/CeO_2$ interfaces under varying levels of activity for CO oxidation. Atomic-resolution imaging of the $Pt/CeO_2$ catalyst is performed during the simultaneous measurement of the catalytic CO oxidation reaction rate. Finite element simulations of the *operando* ETEM reactor are employed to develop and support the chemical reaction rate analysis. Structural dynamics occurring at/near the sites that comprise the metal-support interface are correlated with the catalyst's turnover frequency for CO oxidation. Additionally, the catalyst is observed at 300 °C in an inert atmosphere of $N_2$ to differentiate spectator structural dynamics associated with applied heat from those associated with catalysis. Uniquely, during CO oxidation, the increasing frequency of catalytic turnover is seen to correlate with an increasing concentration of $CeO_2$ surface oxygen vacancies and dynamic structural behavior marking an enhanced rate of oxygen vacancy creation and annihilation; at the same time, the ~1.5 nm Pt nanoparticles become increasingly destabilized and undergo continuous and more frequent fluxional behavior. While the particular fluxional behaviors reported here are specific to the Pt/$CeO_2$ system and to CO oxidation, breaking and forming chemical bonds is an essential functionality for all heterogeneous catalysts, and it may only occur if the catalytic surface undergoes substantial structural dynamics. The *operando* electron microscopy approach described here should be applicable to a large number of nanoparticle catalysts, which will enable the identification of catalytically functional surface structures and strengthen our ability to establish (dynamic) structure-activity relationships.

## 2. Methods

### 2.1. Catalyst Preparation and Characterization of Catalyst Structure and Activity

The catalyst, which consists of Pt nanoparticles supported on nanoparticles of $CeO_2$, was produced by standard hydrothermal (Mai et al., 2005) and incipient wetness impregnation methods, which are described in detail in the **Supplemental Information**. A high weight loading (17 wt.%) of Pt metal was desired to reduce the amount of time spent searching for Pt nanoparticles close to a zone-axis orientation during *in situ* and *operando* TEM experiments. The as-prepared catalyst powder was thermally processed in a flowing stream of 5% $H_2$/Ar for 2 hours at 400 °C prior to any structural or catalytic activity characterization.

Structural characterization was performed using X-ray diffraction (XRD) and aberration-corrected (scanning) TEM ((S)TEM). XRD patterns were collected on a Bruker D-5000 with a Cu Kα source ($\lambda$ = 0.15406 nm). HRTEM images were collected on a Thermo Fisher Titan operated at 300 kV, with the lens system's 3$^{rd}$-order spherical aberration coefficient (-Cs) tuned to approximately -13 µm to yield increased white-column phase contrast (Jia et al., 2010). The STEM (a probe-corrected JEOL ARM200F) was operated at 200 kV and high angle annular dark field (HAADF) images were collected to determine the Pt nanoparticle size distribution, which allowed for an estimate to be made of the number of Pt atoms located at the interfacial perimeter. A derivation of this estimation is given in the **Supplemental Information**.

Catalytic activities for CO oxidation were evaluated in a packed bed plug-flow reactor. Details on the experimental conditions are given in the **Supplemental Information**. Plug flow reactor experiments were performed in a RIG-150 micro-reactor from In Situ Research Instruments (ISRI). Effluent gas compositions were measured with a Varian 3900 gas chromatograph (GC) equipped with a thermal conductivity detector (TCD).

*2.2. Atomic-resolution Operando Environmental TEM*

Atomic-resolution *operando* ETEM experiments were performed on an image-corrected Thermo Fisher Titan ETEM at 300 kV equipped with a Gatan Imaging Filter for electron energy-loss spectroscopy (EELS). An *operando* pellet reactor approach was employed to increase the mass of catalyst in the microscope and facilitate the determination of chemical kinetics during atomic-resolution imaging (Miller et al., 2015; Chenna & Crozier, 2012; Langdon et al., 2019). An overview of the reactor geometry is given in **Figure S4**. In this approach, a Gatan 628 furnace-style Ta hot stage is used for heating. The Ta is inactive. The furnace is loaded with a porous glass fiber pellet and an inert Ta metal TEM grid. Catalyst particles are dispersed on the TEM grid and loaded into the glass fiber pellet through aqueous drop-casting. The mass of catalyst in the pellet was measured with a microbalance and determined to be 180 ± 5 µg. When heated in reactant gases, the large mass of catalyst on the pellet facilitates the production of detectable conversions.

The fiber pellet is annular in shape, having a 1 mm wide hole at its center, which permits the electron beam to pass through unobstructed for spectroscopy and imaging. The TEM grid is visible in the region containing the hole, and it is conducting, so it provides a stable platform for atomic-resolution imaging. The gas composition in the cell can be quantified with EELS (Crozier & Chenna, 2011).

After inserting the catalyst-loaded furnace reactor into the ETEM, an *in situ* reduction step was performed in 1 Torr of $H_2$ at 400 °C for 2 hours. The stage was then cooled to 120 °C and the column was briefly evacuated. As is common for CO oxidation, an oxygen rich reactant environment was used; in this experiment, 0.24 Torr of CO and 0.33 Torr of $O_2$ was admitted into the cell and the reactor was heated up to 300 °C. The pressure of gases used in the ETEM should not significantly impact the kinetic catalytic mechanism given the well-known $0^{th}$-order dependence of the rate on the CO and $O_2$ partial pressures (Cargnello et al., 2013; Liu et al., 2014). Carrier gases were not used to increase the relative partial pressures of the reactants and to achieve high signal-to-background levels for EELS gas composition measurements.

*Operando* TEM images were acquired in the aforementioned negative Cs imaging mode with an incident electron dose rate of $\sim 1 \times 10^3$ e$^-$/Å$^2$/s and an exposure time of 0.5 sec on a Gatan Ultrascan 1000 CCD detector. Previous ETEM work investigating the effect of the incident electron beam dose rate on the structure and chemistry of $CeO_2$ surfaces has demonstrated that electron beam effects are not significant for imaging current densities below $6 \times 10^3$ e$^-$/Å$^2$/s (Sinclair et al., 2017). Additionally, Sinclair et al. (2017) report that no observable structural or chemical changes occur for this dose when low pressures of ambient oxygen are present. Here, oxygen is present, an electron flux below the reported critical threshold is used, and the dose rate is maintained below this level throughout the experiment. Furthermore, work by Lawrence et al.

(2021) shows that, for the imaging conditions employed in the present work, the rate of electron beam-induced radiolytic or knock-on events becomes negligible compared to that of thermally-induced events (i.e., those expected to regulate catalytic processes) as the temperature of the $CeO_2$ rises above 150 °C, which is what is explored here (Ethan L. Lawrence et al., 2021).

The use of a low electron dose rate resulted in a poor signal-to-noise ratio (SNR) in the 0.5 second exposure images that were acquired (for example, see one of frames shown below in **Figures 2b – 2e**). Utilizing a longer image acquisition time was impractical due to instabilities associated with drift in the furnace hot stage required for *operando* TEM. Thus, a drift-corrected, time-averaged image method was implemented to increase the SNR available for local structural analysis under *operando* conditions. A schematic and description of the approach is provided in **Figure S6**.

*In situ* EELS was implemented to track the gas composition (Miller & Crozier, 2014; Crozier & Chenna, 2011; Chenna & Crozier, 2012). All spectra were acquired in diffraction mode using an entrance aperture of 2 mm, a camera length of 245 mm, a dispersion of 0.05 eV/pixel, and an acquisition time of 5 sec. Given the 1:1 stoichiometry of C in the CO oxidation reaction, the *in situ* CO conversion was calculated to be the fractional amount of $CO_2$ to CO and $CO_2$ in the cell.

*2.3. Finite Element Simulation of Operando ETEM Reactor for Chemical Kinetic Analysis*

Recently, we have developed a finite element model of the ETEM reactor, allowing for an investigation into how the reactant conversion measured along the EELS line can be used to derive the steady-state reaction rate for catalyst supported on the Ta TEM grid (Vincent et al., 2020). Here we have employed the model to simulate the *operando* ETEM reactor under conditions matching the present experiment in order to establish a framework for chemical kinetic analysis. The catalytic reaction was modeled as $0^{th}$ order, with kinetic parameters describing the activity of the catalyst taken from the Arrhenius analysis from the plug flow reactor data (**Figure S3**). The spatial distribution of catalyst mass in the pellet was modeled with an egg-shell profile. Simulations were performed in the commercial program COMSOL Multiphysics®. More information describing the simulations is given in (Vincent et al., 2020) and in the **SI.**

*2.4. In Situ ETEM in an Inert Gas Atmosphere at Elevated Temperature*

*In situ* ETEM experiments were conducted on the Pt/CeO$_2$ catalyst in an atmosphere of inert N$_2$ both at room temperature and at 300 °C to differentiate spectator structural dynamics associated with applied heat from those associated with catalysis. These experiments were conducted in an image-corrected Thermo Fisher Titan ETEM operated at 300 kV. TEM samples were prepared by dispersing the Pt/CeO$_2$ powder onto a windowed micro electro-mechanical system (MEMS)-based SiN$_x$ chip, which was then loaded into a DENSsolutions Wildfire heating holder. After loading the holder into the ETEM, 5 mTorr of N$_2$ gas was introduced into the environmental cell, and the catalyst was imaged at room temperature. The specimen was then heated to 300 °C and the catalyst was imaged again. Images were acquired with an incident electron beam dose rate of 5 x 10$^3$ e$^-$/Å$^2$/s using a Gatan K2 IS direct detector in the electron counting mode. The beam was blanked during the *in situ* heating and whenever images were not being acquired to further minimize electron beam induced changes.

## 3. Results

*3.1. Catalyst Morphology and Ex Situ Structural Characterization*

The catalyst's activity for CO oxidation was evaluated in a packed-bed plug-flow reactor. The light-off curves for the bare and Pt-loaded CeO$_2$ are shown in **Figure 1a**. An identical mass loading of catalyst was used for both cases. The activity of the bare CeO$_2$ support (blue) is demonstrably less than the Pt-loaded $CeO_2$ (red) and the effect of blank reactor (black) is negligible at working temperatures. No difference in the $CeO_2$-supported Pt catalyst's activity was detected on the temperature ramp up to full conversion (filled red triangles pointing up) in comparison to the ramp back down to room temperature (empty red triangles pointing down). An Arrhenius analysis of the light-off curves for the Pt/$CeO_2$ catalyst (**Supplemental Figure S3**) shows linear behavior, with an apparent activation energy of $E_{a, app}$ = 74 kJ mol$^{-1}$ calculated from the linear regression, which is consistent with activation energy values reported previously (Cargnello et al., 2013).

The powder XRD patterns of the bare $CeO_2$ and Pt-loaded $CeO_2$ are shown in **Supplemental Figure S1**. A simulated XRD pattern of an infinite crystal of $CeO_2$ (*JCPDS* No. 34-0394) matches well with the experimental XRD patterns of the catalyst, indicating that the sample is phase-pure $CeO_2$ (space group Fm-3m, a = 5.41 Å) along with highly dispersed, small Pt nanoparticles.

A HAADF-STEM image of a typical Pt-loaded $CeO_2$ nanoparticle is shown in **Figure 1b**. The Pt nanoparticles are well-dispersed on the $CeO_2$ nanoparticle support and are around 2 nm in size. The size distribution of the Pt nanoparticles was further investigated and quantified with HAADF-STEM imaging. Details are provided in **Figure S2** of the Supplemental Information. Many different regions of the catalyst sample were imaged and the diameter sizes of 475 Pt nanoparticles were measured, revealing that the average Pt nanoparticle size was 1.6 nm. Quantifying the Pt nanoparticle size allowed for catalytic turnover frequencies (TOFs) to be determined on an interfacial perimeter-site basis. A derivation and sample calculation are provided in the **Supplemental Information**.

**Figure 1c** displays an HRTEM image of a typical $CeO_2$-supported Pt nanoparticle. The Pt nanoparticles are stable and well-facetted when imaged *ex situ*, with the one shown in **Figure 1c** exposing nanosized (100) and (111) surfaces, which is typically observed. The Pt nanoparticles exhibit a well-defined epitaxy with the support, with virtually all nanoparticles presenting (111) surfaces to the metal-support interface, which in this case is comprised of a (111) $CeO_2$ surface on the support side. Well-defined orientation relationships have been reported for $CeO_2$-supported Pt nanoparticles, and they can be attributed to epitaxial relationships that indicate strong structural interactions between the metal and the support (Bernal et al., 2000, 1997; Trovarelli, 2002). The Pt particles remained stable during the HRTEM observation at room temperature.

*3.2. In Situ ETEM Imaging under CO Oxidation Reaction Conditions*

The catalyst was imaged *in situ* under CO oxidation reaction conditions. **Figure 2** shows an *in situ* ETEM image time-series of a typical $CeO_2$-supported Pt nanoparticle in 0.57 Torr of CO and $O_2$ at 144 °C. The Pt particle occupies a short $CeO_2$ (111) nanofacet and each side of the particle is in contact with a $CeO_2$ (111) step edge. Images of the catalyst obtained *in situ* under reaction conditions show prominent motion artifacts and features attributable to fluxional behavior. The time-series is comprised of four images acquired in succession, each with a 0.5 second exposure time. **Figure 2a** displays the time-averaged image obtained by summing together the individual frames over the entire [0 – 2.0] second acquisition period. Relative to the bulk of the underlying $CeO_2$ support, which appears with stable and well-defined white atomic column contrast, the Pt nanoparticle presents white as well as black atomic column contrast. The weak and varying Pt contrast in the 2.0 second time-averaged image of **Figure 2a** is not a consequence of poor microscope imaging conditions, as evidenced by comparison with the bulk $CeO_2$. Rather, the Pt particle is demonstrating fluxional behavior, in which the nanoparticle dynamically progresses through a series of reconfigurations over the course of the time-averaging period. Multislice TEM image simulations have been performed to investigate this effect (see **Supplemental Figure S7**). The simulations demonstrate the appearance of mixed black and white Pt atomic column contrast when the particle is aligned with the incident beam and imaged at an electron optical defocus of 2 nm. Furthermore, the simulations show that contrast reversals can occur when the Pt nanoparticle tilts by a few degrees, e.g., due to a rigid-body rotation.

For the sake of clarity in the following discussion, we define three locations of reference, namely (1) the free $CeO_2$ surface, which is comprised of the sites at the freely exposed $CeO_2$ surface around the Pt nanoparticle, (2) the three-phase boundary, which is comprised of the sites at the perimeter of the metal-support interface, and lastly (3) the buried interface, which is comprised of the sites within the metal-support interface that are not exposed to the gas phase. In the 2.0 sec time-averaged image (**Figure 2a)**, diffuse contrast is observed at the three-phase boundary (right arrow) and along the free $CeO_2$ surface on the $CeO_2$ (111) terrace to the left of the Pt nanoparticle (left arrow). In comparison, the sub-surface and bulk Ce sites show sharp and clearly resolved atomic columns. The diffuse contrast or local streaking/blurring observed at the $CeO_2$ surface and $Pt/CeO_2$ interface is therefore not due to drift in the hot stage or the electron optics, but instead arises from dynamic structural reconfigurations occurring at these specific sites (Ethan L. Lawrence et al., 2021).

An examination of the *in situ* image time-series reveals the atomic-scale structural dynamics and how they evolve over time. **Figures 2b – 2e** show the four individual 0.5 second exposure frames taken over the [0 – 2.0] second acquisition period (the images have been processed with a bandpass filter for clarity). **Figures 2(f1) – 2(f4)** display the Fourier transform (FT) at each time interval, taken from the windowed region around the Pt nanoparticle that is indicated by the dashed box in **Figure 2b**. An analysis of the major spots that appear in the FTs indicate the presence of metallic Pt (see **Supplemental Figure S8**). (The FTs were produced from unfiltered images that were pre-processed with a 2D Hanning function to suppress edge artifacts caused by windowing).

As seen both in the time-series images and corresponding FTs, the Pt nanoparticle undergoes a series of structural reconfigurations that involve dynamic restructuring at the Pt particle surface and at the Pt/CeO$_2$ interface. At the same time, certain surface and interfacial Ce sites exhibit dynamic events that occur along with those observed in the supported Pt. In **Figure 2c**, we observe the catalyst at a moment of relative structural stability, evidenced by the surface faceting and atomic column contrast in the Pt. In **Figure 2c**, the arrowed Ce atomic columns at/near the three-phase boundary appear with diffuse contrast but are clearly resolved. In the next frame, 0.5 seconds later (**Figure 2d**), the same arrowed Ce sites have either disappeared or substantially blurred. The Pt has undergone a rotation or tilt, evidenced by the diagonal streaking in the Pt and by the loss of associated Bragg spots in the FT shown in **Figure 2(f3)**. Another 0.5 seconds later (**Figure 2e**), the Pt particle shows weak contrast with no evident lattice fringes. The particle has also adopted an apparently rounded shape, which is marked with a dashed line in the image. The lack of lattice fringe contrast in the Pt signifies that the supported particle is undergoing dynamic restructuring at a speed that is much faster than the frame exposure time. Interestingly, in this time interval, the Ce sites at the free CeO$_2$ surface and at the three-phase boundary also appear very dynamic. As seen in **Figure 2e**, a build-up of bright but diffuse intensity appears at the perimeter of the Pt/CeO$_2$ interface, which is indicated with the arrows. Neighboring Ce sites now show vacancies, so the bright and diffuse contrast at the interfacial perimeter may indicate that nearby surface atoms have migrated to the perimeter sites, where they continue to undergo rapid dynamics that cannot be resolved with the present frame rate (see, e.g. (Crozier et al., 2019)).

At this condition, the reactor temperature is below the light-off point of the catalyst, and no conversion is detected yet in the ETEM cell. While the ensemble of catalyst particles loaded in the ETEM reactor is not yet active enough to produce a measurable conversion, the Pt/CeO$_2$ catalyst still undergoes a variety of dynamic structural reconfigurations which constantly evolve over time. Extrapolating the reaction rates experimentally measured at higher temperatures (and presented in **Figure 3**) to this temperature, 144 °C, results in a turnover frequency on the order of $5 \times 10^{-5}$ molecules CO site$^{-1}$ sec$^{-1}$, showing that continuous catalytic turnover is not occurring at an appreciable rate. As such, the dynamic structural reconfigurations observed in **Figure 2** are associated with intermediate steps in the catalytic reaction that have a lower activation energy than the rate-limiting step. Recent density functional theory calculations indicate that abstraction of lattice oxygen by CO on $CeO_2$ (111) surfaces proceeds with an activation energy barrier around 0.4 eV (Salcedo & Irigoyen, 2020).

On the other hand, it is generally well known that the molecular dissociation and gas phase exchange of $O_2$ is energetically much more challenging, with theoretical calculations and experimental isotopic oxygen exchange measurements placing the activation energy over $CeO_2$ around 1.1 eV (Martin & Duprez, 1996; Han et al., 2019). Hence, here we speculate that the observed fluxional behavior is associated with the energetically more facile steps of lattice oxygen abstraction by CO, and that the absence of measurable conversion is due to the lack of oxygen vacancy back-filling from molecular $O_2$ exchange. This fluxional behavior was not observed at room temperature in vacuum. Additional imaging experiments carried out in a $N_2$ atmosphere at a temperature of 300 °C show a lack of similar structural dynamics, further demonstrating that the fluxional behavior is not due simply to thermal effects (see **SI Section 7**).

Finally, it is worth noting that the dynamic behavior exhibited by the $CeO_2$-supported Pt nanoparticle in **Figure 2** is typical of the many observed during the ETEM experiment. Additional images and analysis from a different particle, for example, are shown in **Supplemental Figure S12**. We have also carried out *in situ* TEM imaging of $CeO_2$-supported Pt nanoparticle catalysts in atmospheres of CO (Crozier et al., 2019), CO and $O_2$ (Vincent & Crozier, 2020), and CO and $H_2O$ (Li et al., 2021), and shown that substantial fluxional behavior can occur in these reactant gases – even at room temperature. In order to identify structural dynamics unambiguously associated with catalytic chemistry, we proceed to image the catalyst under *operando* conditions and correlate the observed fluxional behavior directly with the chemical kinetics measured in the ETEM cell.

*3.3. Detection and Quantification of Catalytic Reaction Rate in the ETEM Reactor*

The catalyst was heated until detectable conversions of CO into $CO_2$ occurred. The catalytic reaction rate in the ETEM reactor was determined by experimental EELS measurements supported by finite element simulations. **Figure 3a** displays a set of background-subtracted and normalized energy-loss spectra to demonstrate the detection of catalytically-produced $CO_2$. When no catalyst is loaded in the reactor, the spectrum of gas around the reactor (dashed gray curve) shows a prominent C $\pi^*$ peak at 285 eV, which corresponds to CO. The conversion of CO to $CO_2$ is observed only after $Pt/CeO_2$ catalyst is present (solid red curve). In this case, a second prominent peak appears at 288.3 eV which corresponds to the C $\pi^*$ peak of $CO_2$. Both spectra were acquired under nominally identical conditions for a furnace set point of 300 °C in 0.57 Torr of CO and $O_2$.

The composition of CO and $CO_2$ present in the spectrum can be quantified to calculate the CO conversion (Miller & Crozier, 2014; Crozier & Chenna, 2011; Chenna & Crozier, 2012). **Figure 3b** presents the EELS CO conversions detected during the *operando* ETEM experiment, plotted as a function of the reactor temperature (red circles). At 251 °C, the CO conversion was calculated to be 2.8 ± 3.7%, increasing at 275 °C to a value of 15.1 ± 0.5% and further at 285 °C to a value of 19.8 ± 0.3%. Finite element simulations of the ETEM reactor were performed to establish a framework for linking the CO conversion measured along the EELS line to the reaction rate of the catalyst imaged on the TEM grid. Further details on the simulations are given in the **Supplemental Information** and in (Vincent et al., 2020). Steady-state calculations were done under conditions nominally identical to the current experiment, and the gas and temperature distributions were determined during catalysis. The simulated CO conversion measured along the EELS line in the model is plotted as the solid gray curve in **Figure 3b**. The simulated conversion curve shows reasonable agreement with the values measured experimentally.

It is important to investigate the extent of any thermal gradients that may exist within the reactor. **Figure 3c** shows the simulated 3-dimensional temperature distribution in the *operando* ETEM reactor for a furnace thermocouple set point of 275 °C. The temperature distribution appears largely uniform throughout the reactor, and the temperature of the reactor matches well with the furnace set point. The temperature distribution shown in **Figure 3c** is representative of the behavior observed at other working temperatures (see, e.g., **Supplemental Figure S5**). The high degree of thermal uniformity suggests that the reactor can be treated as approximately isothermal. A quantitative comparison of the temperature difference between the furnace set point and the TEM grid (see **Table S2**), reveals that the typical discrepancy between these two locations is at most 1.6 °C, supporting this assumption.

Next we investigate how the EELS CO conversion measurement can be used to derive quantitative information about the reaction rate of catalyst particles on the TEM grid. Experimentally, the catalytic reaction rate (i.e., the rate of product formation, $r_{CO_2}$, with units of mol $CO_2$ per second) may be estimated from the EELS CO conversion measurement, $X_{CO}$, by multiplying the CO conversion with the known inlet molar flow rate of CO, $\dot{n}_{CO,in}$:

(1) $r_{CO_2} = \dfrac{X_{CO} \times \dot{n}_{CO,in}}{m_{cat}}$

A common convention in the catalysis literature is to normalize the catalytic reaction rate by the mass of catalyst in the reactor, $m_{cat}$. The mass-normalized $r_{CO_2}$ estimated from the EELS CO conversion (i.e., Equation 1) may also be simulated in the model, since it is possible to replicate the conversion measurement by integrating the gas composition along the electron beam path (see **SI** or (Vincent et al., 2020) for an extended discussion). The simulated $r_{CO_2}$ estimated from the EELS CO conversion is plotted as a function of temperature as the dashed red curve in **Figure 3d**.

The rate estimated from the conversion (i.e., Equation 1) should not be taken as the rate of catalyst particles on the TEM grid. Previously we have shown that the difference between the estimated rate and the true rate at the grid can vary significantly as a function of conversion and grow beyond 200% in some cases, clearly demonstrating the need for and power of a model in determining quantitative chemical kinetics. In the model, the true rate of product formation may be found by integrating the reaction rate throughout the domain of the pellet. The rate may be normalized to mass by also integrating the mass distribution within the same domain. As explained in the **SI** and in Vincent et al. (2020), the mass-normalized rate for the particles on the TEM grid can be simulated by integrating the mass and rate along the innermost surface of the pellet, where the composition and temperature are both nearly identical to that of the TEM grid (Vincent et al., 2020). This quantity is plotted as a function of temperature as the solid blue curve in **Figure 3d**. Observe that the estimated and true rate are not equivalent, which underscores the importance of establishing a model in order to relate the *in situ* conversion measurement to the true activity of the imaged catalyst for quantitative *operando* TEM. An analysis of the ratio of the two rates (see **Supplemental Table S1**) provides a conversion-dependent correction factor that allows the true rate to be calculated correctly from the rate estimated through the EELS conversion measurement.

Finally, we choose to normalize the reaction rate on an interfacial perimeter site-basis, yielding an ensemble *in situ* TOF. As discussed, a relationship has been derived to link catalyst mass with the average number of Pt perimeter sites (see **SI**). Thus, the corrected *in situ* reaction rate can be used to estimate an *in situ* TOF for catalyst particles on the TEM grid. This quantity is plotted along the right axis of **Figure 3b**. Note that the plotted TOF values were obtained after correcting the rate measured with EELS according to the finite element analysis presented above. At 251 °C, the *in situ* TOF was calculated to be 0.15 CO site$^{-1}$ sec$^{-1}$, increasing at 275 °C to a value of 0.80 CO site$^{-1}$ sec$^{-1}$ and further at 285 °C to a value of 1.05 CO site$^{-1}$ sec$^{-1}$.

*3.4. Fluxional Dynamics Occurring at and near the Pt/CeO$_2$ Interface During Catalysis*

With a quantitative measure of the chemical reaction rate, we proceed to image the catalyst at varying degrees of activity and then correlate the measured reaction kinetics directly with atomic-resolution observations of the dynamic working catalyst structure. **Figure 4** displays a set of three 12.5 second time-averaged *operando* ETEM images of the same CeO$_2$-supported Pt nanoparticles taken at (a) 144 °C, where the TOF was determined to be 0.00 CO site$^{-1}$ sec$^{-1}$, (b) at 275 °C, where the TOF was measured to be 0.80 CO site$^{-1}$ sec$^{-1}$, and (c) at 285 °C, where the TOF was measured to be 1.05 CO site$^{-1}$ sec$^{-1}$. The roughly 1.5 nm Pt nanoparticles are supported on small ~2 nm CeO$_2$ (111) nanofacets, with either side of each nanoparticle in contact with CeO$_2$ (111) step edges.

The increasing frequency of catalytic turnover is seen to coincide with dynamic fluxional behavior in the working catalyst structure. The observed dynamic fluxional behaviors give rise to motion artifacts and contrast features in the time-averaged *operando* TEM images that are similar to those discussed previously in **Section 3.2**. Here we observe that the Pt nanoparticles evolve from an initially well-defined, facetted morphology, with clearly visible lattice fringe contrast (**Figure 4a**), to darker patches of nearly uniform contrast, now exhibiting an apparently rounded morphology, with lattice fringes that are greatly diminished in visibility (**Figure 4c**). Additionally, the Ce sites on the free $CeO_2$ surface neighboring the metal-support interface become increasingly more blurred with larger catalytic turnover. Finally, in addition to a pronounced blurring at higher turnover, the top layer of Ce that neighbors the $Pt/CeO_2$ interface also displays an outward surface relaxation which grows larger with increasing activity. The outward surface relaxation is difficult to see by eye in the image but is easily measurable in intensity line profiles taken at each condition (discussed below and presented in **Supplemental Figure S15** and **Table S3**).

A quantitative analysis of the *operando* ETEM images was performed in order to establish correlations between these fluxional dynamics and the catalyst's turnover frequency for CO oxidation. To describe the degree of structural dynamics taking place in the Pt nanoparticles, we quantify the Pt lattice fringe visibility (**Figure 5**) through FT analysis. Windowed regions centered on the left and right Pt NPs were extracted from the time-averaged *operando* ETEM images at each condition, then filtered with a Hanning function, and finally transformed into Fourier space. The left and right Pt NP FT window are shown, at 144 °C for example, in the left-most side of **Figure 5**. **Figures 5(a1)-(a3) and 5(b1)-(b3)** show the modulus of the FTs taken from the right and left Pt NP, respectively, at conditions corresponding, respectively, to catalytic TOFs of 0.00, 0.80, and 1.05 CO site$^{-1}$ sec$^{-1}$.

The Pt fringe visibility clearly decreases with increasing catalytic turnover. In **Figure 5(a1)**, labels are given to Bragg spots corresponding to various low-index Pt fringes. The Pt lattice fringe visibility can be quantified by summing the magnitude of the Bragg spots corresponding to Pt (here we sum the FT modulus within the regions defined by the dashed yellow circles as shown in **Figure 5(a1)**; the image intensities were normalized prior to summation). We call this quantity the Pt FT component strength and plot it for the left and right Pt NP as a function of catalytic turnover frequency in **Figure 5c**. The component strength of the left (green triangles) and right (blue squares) Pt NP are similar at each temperature. As the turnover frequency for CO oxidation increases, the visibility of fringes in either NP decreases. The component strengths of the left and right NP are 40.6 and 46.3, respectively, at 144 °C when the TOF is 0 CO site$^{-1}$ sec$^{-1}$. At 275 °C, the TOF increases to 0.80 CO site$^{-1}$ sec$^{-1}$ and the left and right Pt NP component energies drop to 30.6 and 33.4, respectively. At 285 °C, where the TOF has increased to 1.05 CO site$^{-1}$ sec$^{-1}$, the left and right Pt NP component energies decrease again to 25.1 and 28.6, respectively. As seen in the time-averaged *operando* TEM images (**Figure 3c**) and in the associated FTs (**Figure 5(a3) and 5(b3)**), at this condition, the lattice fringes are diffuse in either Pt NP and Bragg spots are significantly weaker.

Next we focus on the fluxional behavior and structural dynamics occurring on the free (111) $CeO_2$ surface that connects the two Pt particles. Enhanced views of the connecting terrace are given in **Figure 6** for TOFs of (a1) 0.00 CO site$^{-1}$ sec$^{-1}$, (a2) 0.80 CO site$^{-1}$ sec$^{-1}$, and (a3) 1.05 CO site$^{-1}$ sec$^{-1}$. The surface Ce sites neighboring the metal-support interface become increasingly blurred with higher catalytic turnover. As can be seen in Figure 6(a1), the free $CeO_2$ surface sites neighboring the Pt/$CeO_2$ interface already shown a visible degree of blurred intensity at a TOF of 0.00 CO site$^{-1}$ sec$^{-1}$. However, when the TOF increases to 1.05 CO site$^{-1}$ sec$^{-1}$, the blurring of the surface Ce becomes further pronounced to the point that the top layer appears as a nearly continuous band of intensity running parallel to the (111) $CeO_2$ support surface (**Figure 6(a3)**). A similar blurring of the free surface Ce sites located at the opposite sides of the Pt nanoparticles can be observed in **Figure 4c**. The blurring does not appear within the $CeO_2$ bulk, indicating that the contrast change in the image is a consequence of fluxional changes in the $CeO_2$ surface near the metal-support interface.

The degree of blurring can be quantified and correlated with TOF by computing the average free surface Ce atomic column contrast at each condition. Here, the atomic column contrast is calculated by analyzing 100 pm wide intensity line profiles taken over the region that contains the surface Ce cation columns. A description of the approach along with the intensity line profiles used in the analysis is given in **Figure S12**. The average surface Ce atomic column contrast is plotted as a function of catalytic turnover frequency in **Figure 6b**; error bars are given as the standard deviation relative to the mean measurement.

The contrast measurements plotted in **Figure 6b** show unambiguously that the fluxional behavior giving rise to the Ce column blurring becomes more pronounced at higher catalytic turnover. At 144 °C when the TOF is 0 CO site$^{-1}$ sec$^{-1}$, the average surface Ce contrast is 0.031. At 275 °C, the TOF has increased to 0.80 CO site$^{-1}$ sec$^{-1}$, and the average contrast has decreased to a value of 0.017. At 285 °C the TOF has again increased to 1.05 CO site$^{-1}$ sec$^{-1}$, although the average measured contrast remains more or less the same at a value of 0.018. Upon further investigation, it became clear that the contrast measurement at 275 °C was biased toward lower values due to the fact that the support particle had tilted, leading to an enhanced streaking in the direction of the measurement. An examination of intensity line profiles taken from the bulk of the CeO$_2$ show an analogous reduction in contrast (**Figure S13b**). As the catalyst was heated up to 285 °C, though, the crystal tilted back to roughly the same orientation, evidenced by the recovered contrast from columns in the bulk of the nanoparticle as seen in **Figure S13b**. Since the CeO$_2$ support has returned to roughly the same [110] zone axis orientation at 285 °C, comparing the quantified contrast of the columns at 144 °C vs 285 °C can provide insight into the evolution of blurring that occurs at individual surface Ce sites near the metal-support interface. **Figure S14** presents the quantified contrast of individual surface Ce columns at a TOF of 1.05 CO site$^{-1}$ sec$^{-1}$ plotted against the quantified contrast of the same column when the TOF was measured to be 0 CO site$^{-1}$ sec$^{-1}$. The plot demonstrates that while on average the contrast decreases with increasing turnover, one column (located in the middle of the surface row) actually shows a slight increase in contrast, which indicates that redistributions of cation column occupancy along the surface are occurring as well as the dynamic behavior, which gives rise to the observed blurring.

In addition to a pronounced blurring at higher turnover, the top layer of Ce that neighbors the Pt/CeO$_2$ interface also displays an outward surface relaxation which grows larger with increasing activity. The outward CeO$_2$ (111) surface relaxation was quantified by measuring the lattice plane separation distance from intensity line profiles taken from the interior of the CeO$_2$ support toward the catalyst surface. A description of the approach along with the intensity line profiles used in the analysis is provided in **Figure S15**. The separation distance measured between the surface and subsurface CeO$_2$ (111) lattice planes is plotted as a function of catalytic turnover frequency in **Figure 6c**. The error bars are derived from the standard deviation about the mean lattice spacings measured in the subsurface and bulk at that condition (here, we briefly note that the average of the sub-surface and bulk lattice plane separation distances was measured to be 310 ± 4 pm, which agrees with the accepted CeO$_2$ (111) Miller plane spacing of 312 pm).

Examining the surface lattice spacing plotted against catalytic turnover frequency illustrates the emergence of an outward surface relaxation that grows linearly with increasing activity. At 144 °C where the TOF is 0 CO site$^{-1}$ sec$^{-1}$, the surface CeO$_2$ (111) lattice plane distance is 306 ± 4 pm. At 275 °C, where the TOF has increased to 0.80 CO site$^{-1}$ sec$^{-1}$, the surface CeO$_2$ (111) lattice plane has relaxed outward by 9 pm to a value of 315 ± 3 pm. At 285 °C, the TOF has increased to 1.05 CO site$^{-1}$ sec$^{-1}$, and the surface CeO$_2$ (111) lattice plane has relaxed by another 5 pm to a value of 320 ± 5 pm. The surface lattice plane separation was also measured in an inert atmosphere of $N_2$ at room temperature and at working temperature; no such expansion was detected upon heating the catalyst up to 300 °C, as shown in **Figure S11**.

Considering the substantial dynamic structural behavior as well as the outward $CeO_2$ support surface relaxation observed to occur during catalysis, one may suspect significant local lattice strain to be present. We have measured the strain in the $CeO_2$ by fitting 2D elliptical Gaussians to the Ce columns in the time averaged image (Levin et al., 2020), in order to determine their position; then we have calculated the difference in each columns' position with respect to the bulk terminated lattice, in a standard way. We call this quantity the *average static strain* since it is measured from the time-averaged images, which are the sum of image signals recorded over a total time period of 12.5 s. The in-plane component of the average static strain is plotted in **Figure 7** for TOFs of (a) 0.00 CO site$^{-1}$ sec$^{-1}$ and (b) 1.05 CO site$^{-1}$ sec$^{-1}$ (strain analysis for data corresponding a TOF of 0.85 CO site$^{-1}$ sec$^{-1}$ was not performed due to the aforementioned crystal tilt at this condition). In the figure, the fitted positions of Ce atomic columns are represented by circles. The in-plane component of the average static strain shows rapid spatial variation between tensile and compressive across the free $CeO_2$ surface terrace, with a magnitude in the range of 15 – 20%.

While the static strain at the surface is large, it does not appear to show any clear correlation with the TOF. We believe this is because the average static strain may not give an accurate representation of the dynamic bond distortion that may be present on the surface at any given time due to fluxional behavior. For example, if during an image exposure time, *t*, the cation spends a time *t*/2 at distance +*d* away from its bulk terminated position and a time *t*/2 at a distance -*d*, then the average cation displacement over the observation time would be zero. To address this deficiency, we define a *fluxional strain* quantity that is proportional to the width of the atomic column in the time-averaged image. The fluxional strain, $\varepsilon_{flux}$, can then be taken as the difference in the standard deviation of the Gaussian fitted to the particular Ce cation column of interest, $\sigma_{local}$, and in the bulk, $\sigma_{bulk}$, and is defined in a manner similar to the static strain, as follows:

(2) $\quad \varepsilon_{flux} = \frac{(\sigma_{local} - \sigma_{bulk})}{\sigma_{bulk}}$

The local magnitude of the fluxional strain is visually represented in **Figure 7** by the size of the circles, which are proportional in dimension to the standard deviation of the Gaussian fitted to the Ce atomic columns. Note that the widths of the atomic columns in the bulk of the $CeO_2$ do not vary significantly with increasing turnover frequency. The columns at the surface, however, grow wider with increasing turnover, as can be seen by examining the circles drawn at the top of the figures in the arrowed locations, which correspond to the atomic columns located at the free $CeO_2$ (111) surface terrace. The average fluxional strain across the surface terrace is around 19% at 144 °C when the conversion is zero. However, as the conversion increases, the fluxional strain also increases and is around 33% at 285 °C when the conversion is 15%. This is a large degree of strain, and the resultant high surface energy will make the catalyst surface extremely reactive.

## 4. Discussion

The observations and measurements clearly show significant dynamic changes in the metal particles, the metal/ceramic interface, and the nearby support surfaces under reaction conditions and during catalysis. The rate of structural dynamics at or near the three-phase boundary correlates with the TOF and thus provides insights into the atomic level materials processes taking place during a Mars van Krevelen catalytic reaction. Interestingly, the structural changes are observed to some degree even when the TOF is zero. However, as we will discuss below, it would be an oversimplification to conclude that these are spectator processes, since they are associated with necessary intermediate reactions steps even when the rate limiting step is not completed.

Moreover, the catalyst changes form as the TOF increases due to the substantial increases in fluxional behavior. The fact that the changes become more pronounced when catalysis is detected suggests that they are integral to the functioning of the Mars van Krevelen mechanism in this case. The various dynamic processes are interconnected, but it is helpful to initially discuss them separately.

The most obvious change observed in our experiments is the emergence of structural dynamics taking place in the Pt particle under reaction conditions. Significant fluxional behavior of the Pt nanoparticles takes place even at 144 °C, which is below the light-off temperature for the catalyst. We also observed similar behavior in recent *in situ* TEM on the water gas shift reaction at 200 °C (Li et al., 2021) and we have even observed significant fluxional behavior in Pt particles supported on $CeO_2$ at room temperature in atmospheres containing CO (Crozier et al., 2019) (but not in $N_2$, for example see **Figures S9/S10**). It is clear that this behavior is not directly associated with the rate limiting step for CO oxidation, but one may conclude that the equilibrium shape of 1 – 2 nm Pt nanoparticles supported on $CeO_2$ is not well-defined in a CO atmosphere over a wide range of temperatures.

Under equilibrium conditions, a supported nanoparticle will adopt the so-called Winterbottom shape (Winterbottom, 1967) which minimizes the sum of the surface and interfacial energies. For an FCC metal like Pt, the (111) surface has the lowest energy in vacuum (McCrum et al., 2017) and the coherent interfaces associated with strong bonding (e.g., the one shown in **Figure 1**) are also low energy. In this case, the strong interfacial bond between Pt and $CeO_2$ is associated with bridging oxygen giving rise to Pt – O – Ce linkages (Hatanaka et al., 2010; Shinjoh et al., 2009; Nagai et al., 2006; Gänzler et al., 2017, 2018). For these metallic systems, the Winterbottom shape often gives rise to a truncated Wulff shape (the equilibrium shape of an unsupported particle). In the presence of gas adsorbates, the surface energies may change, causing the equilibrium particle shape to change (Hansen et al., 2002; Zhu et al., 2020). As is well known, CO binds very strongly to Pt with an average chemisorption energy of about 1.3 – 1.5 eV (Lu et al., 2020; Podkolzin et al., 2000). It is important to recognize that the formation of this strong Pt-CO bond will also weaken the Pt bond with its nearest neighbors. Under reaction conditions, most surface Pt atoms will be bonded to CO (Podkolzin et al., 2000; Lu et al., 2020), and so there will be a substantial weakening of the bond between the surface and the subsurface Pt atoms.

Examination of the Ce cations at the Pt perimeter sites in **Figure 4a** shows that oxygen vacancy creation and annihilation is taking place due to interaction with CO adsorbed on the Pt. For 2 nm Pt particles, assuming a simplified hemispherical cuboctahedral shape, approximately 35% of the Pt atoms are located at the Pt/$CeO_2$ interface and about 40% of those atoms occupy the perimeter sites. Oxygen vacancy creation at these perimeter sites removes bridging oxygen and weakens the bonding between the Pt nanoparticle and support. The fluxional behavior in the surface Ce cations suggests that there is constant competition between vacancy annihilation, which repairs the Pt-O-Ce bond, and vacancy creation, which is due to reaction with CO. We can get additional temporal insights by considering the image exposure time. Over a 0.5 sec exposure at 144 °C (see **Figure 2b-e**), the significant blurring of the perimeter Ce cations indicates that many vacancies are created and annihilated due to interactions with CO.

At this temperature, no $CO_2$ is detected, so the carbonaceous intermediates may remain bound to the catalyst's surface (indeed, several authors report that $CO_2$ adsorbed on *reduced* $CeO_{2-x}$ binds with an adsorption energy in excess of 1 eV, as discussed by D. R. Mullins in (Mullins, 2015)). Ionic oxygen transport is also small at 144 °C, so there is a limited availability of lattice oxygen to backfill the surface and interfacial vacancies. Thus, much of the vacancy creation/annihilation will be associated with forward and reverse reactions between surface lattice O and CO, *i.e.* this intermediate step may be highly reversible. The continuous disruption of the Pt-O-Ce interfacial bond and the changes in adsorbate configuration on the Pt surface disrupt the Winterbottom shape and leads to the observed dynamic behavior.

As carbon species spillover onto the adjacent $CeO_2$ surface, additional CO will adsorb onto the vacant surface sites on Pt. This dynamic bonding and debonding of adsorbates on the metal surface leads to distortions and strains, driving surface diffusion. In recent work, we correlated *in situ* TEM and *in situ* synchrotron observations. The synchrotron analysis showed that, even when rapid fluxional behavior is occurring, the average Pt coordination and bond lengths are consistent with the Pt FCC crystal structure (Li et al., 2021). This shows that the nanoparticle is not forming an amorphous structure, but rather it is transforming rapidly between metastable configurations. At present, our temporal resolution and signal-to-noise preclude a detailed identification of the structure of these metastable phases. However, the analysis of the visibility of the Bragg beams in the diffractograms shows that the rate of reconfiguration increases with conversion. This is consistent with more frequent formation and breaking of surface and interfacial chemical bonds associated with high catalytic TOF.

As the temperature and conversion increase, there is a substantial increase in the fluxional behavior on the $CeO_2$ surface. While the highest vacancy creation/annihilation activity takes place at the perimeter sites, the fluxional behavior extends over the entire 20 Å nanofacet between the two Pt particles shown in **Figure 6c**. The outward relaxation of the surface layer is also consistent with a high concentration of oxygen vacancies along the surface. No lattice expansion was detected in the subsurface layer, and by the third layer down, no evidence is seen for any Ce cation blurring. This localized surface fluxional behavior indicates that most of the oxygen vacancy creation and annihilation as well as oxygen diffusion takes place on the top layer. Negligible oxygen transport from the bulk should take place at these temperatures since the activation energy undoped $CeO_2$ has been measured to lie in the range 0.9 – 2.3 eV (Wu et al., 2010; Kamiya et al., 2000). At conditions of detectable conversion, this implies that at least some of the oxygen vacancies must be annihilated via reduction of molecular oxygen in order to replenish the supply of oxygen required for steady state oxidation of CO.

One may ask *where is the likely site for molecular oxygen reduction?* It could take place at the oxygen vacancies created at the perimeter sites after $CO_2$ desorbs. Alternatively, it could take place on the nearby free $CeO_2$ terrace and then diffuse to the perimeter sites. As shown in **Figure 7**, the average static surface strain is high on the free $CeO_2$ surface, and the fluxional strain on the surface terrace reaches around 30% when the conversion is 15% – this is a very large strain and the associated high surface energy will make the terrace more reactive making direct oxygen reduction feasible. At present we are not able to differentiate definitively between either perimeter or terrace mechanism, though it may be worth speculating on the likelihood of each. We do not have atomic resolution surface spectroscopy to determine the nature of the carbon species created as oxygen vacancies are created and annihilated. However, carbonates are known to form during CO oxidation on $CeO_2$ (Wu et al., 2012; Mullins, 2015), so a possible reversible reaction that may occur at the interface is:

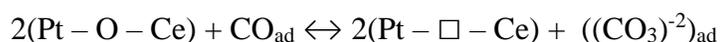

$$2(Pt - O - Ce) + CO_{ad} \leftrightarrow 2(Pt - \square - Ce) + ((CO_3)^{-2})_{ad}$$

The perimeter environment may be rather crowded with a high concentration of CO on the Pt sites and a high concentration of carbonates on the $CeO_2$ sites. Backfilling of the interfacial oxygen vacancies by direct adsorption of molecular oxygen may be subject to significant steric hinderance. It may be easier to annihilate the interfacial oxygen vacancy by diffusion of lattice surface oxygen along the nearby CeO$_2$. During catalysis, oxygen reduction almost certainly takes place on the highly reduced nearby CeO$_2$ terrace and that lattice oxygen migrates to the perimeter sites for reaction with CO.

These experiments provide an atomic level view of the structural dynamics associated with the three-phase boundary of the catalyst during Mars-van Krevelen oxidation. The localized structural changes that are observed are directly associated with catalytic functionality. Many of the steps for converting reactants into products involve forming and breaking chemical bonds with the atoms forming the catalyst surface. This bonding and debonding not only changes the adsorbates but may locally destabilize the catalyst surface and interface structures, resulting in large local surface strains, surface atom migration, and the creation and annihilation of point defects. The particular fluxional behaviors reported here are specific to the catalytic system and CO oxidation. However, breaking and forming chemical bonds is an essential functionality for all heterogeneous catalysts, and it can only occur if the catalytic surface undergoes substantial structural dynamics. While the electron transfer occurring at an active site happens on the femto-second time scale, the subsequent nuclear rearrangements may alter or completely destroy the active site due to the thermodynamic and kinetic factors driving structural change. The question of how to explore structure-reactivity relations then becomes more complex: in at least some nanoparticle catalysts, it may be more appropriate to carefully consider the atomic-level *structural dynamics* that may be integral to the catalytic cycle and reactivity. The importance of adsorbate-induced structural changes has long been recognized in the surface science community (Somorjai, 1991). Now, however, the picometer precision of advanced *operando* TEM can reveal the local, atomic-level details of these structural dynamics for actively working technical catalysts.

## 5. Conclusion

In summary, we have employed aberration-corrected *operando* TEM to visualize the atomic-scale dynamic structural (i.e., fluxional) behavior occurring at and near Pt/CeO$_2$ metal-support interfaces during oxygen transfer, with a focus on CO oxidation. Finite element modeling is performed to develop a reaction rate analysis wherein the atomic-resolution structural observations are directly correlated with the catalyst's turnover frequency for CO oxidation. We show that the increasing frequency of catalytic turnover correlates with dynamic fluxional behavior that (a) destabilizes the supported Pt particle, (b) marks an enhanced rate of oxygen vacancy creation and annihilation, and (c) leads to increased strain and reduction in the surface of the CeO$_2$ support. These results unambiguously demonstrate that there is a dynamic transformation in the chemical environment around the metal-support interface during catalysis. The results also show that the equilibrium shape of 1 – 2 nm Pt nanoparticles supported on CeO$_2$ is not well-defined in a CO atmosphere, especially at high catalytic turnover, where the larger concentration and faster cycling of oxygen vacancies contributes to the destabilization of the supported Pt nanoparticle. Overall, the results implicate the interfacial Pt-O-Ce bonds anchoring the Pt to the support as being involved also in the catalytically-driven oxygen transfer process, and they suggest that molecular oxygen reduction takes place on the highly reduced nearby CeO$_2$ surface before migrating to the metal-support interfacial perimeter for reaction with CO. This study highlights the importance of characterizing the structural dynamics that take place during catalysis in order to elucidate the relationship between a catalyst's structure and its functionality.

## 6. Acknowledgements

The authors gratefully acknowledge funding for this research from NSF grant CBET 1604971. The authors thank Arizona State University's John M. Cowley Center for High Resolution Electron Microscopy for microscope access and use. The authors also gratefully acknowledge use of environmental electron microscopy facilities at the National Institute of Standards and Technology in Gaithersburg, MD, and in particular would like to acknowledge the helpfulness and hospitality of Dr. Wei-Chang David Yang, Dr. Canhui Wang, and Dr. Renu Sharma. Additionally, the authors are grateful to Mr. Piyush Haluai for assistance with measuring the strain of Ce atomic columns.

## 7. Contributions to Research

JLV performed the experiments and analysis. JLV wrote the manuscript with input from PAC. PAC conceived of the research, supervised all aspects of the project, and edited the manuscript.

## 8. Declaration of Competing Interests

The authors declare no competing interests.

## 9. References

ANEGGI, E., BOARO, M., COLUSSI, S., DE LEITENBURG, C. & TROVARELLI, A. (2016). Ceria-Based Materials in Catalysis: Historical Perspective and Future Trends. *Handbook on the Physics and Chemistry of Rare Earths* **50**, 209–242.

BALAJI GOPAL, C., GARCÍA-MELCHOR, M., LEE, S. C., SHI, Y., SHAVORSKIY, A., MONTI, M., GUAN, Z., SINCLAIR, R., BLUHM, H., VOJVODIC, A. & CHUEH, W. C. (2017). Equilibrium oxygen storage capacity of ultrathin $CeO_{2-\delta}$ depends non-monotonically on large biaxial strain. *Nature Communications* **8**, 15360.

BAÑARES, M. A. (2005). Operando methodology: Combination of in situ spectroscopy and simultaneous activity measurements under catalytic reaction conditions. *Catalysis Today* **100**, 71–77.

BERNAL, S., BAKER, R. T., BURROWS, A., CALVINO, J. J., KIELY, C. J., LÓPEZ-CARTES, C., PÉREZ-OMIL, J. A. & RODRÍGUEZ-IZQUIERDO, J. M. (2000). Structure of highly dispersed metals and oxides: Exploring the capabilities of high-resolution electron microscopy. *Surface and Interface Analysis* **29**, 411–421.

BERNAL, S., CALVINO, J. J., GATICA, J. M., LARESE, C., LÓPEZ-CARTES, C. & PÉREZ-OMIL, J. A. (1997). Nanostructural evolution of a Pt/CeO2 catalyst reduced at increasing temperatures (473-1223 K): A HREM study. *Journal of Catalysis* **169**, 510–515.

CARGNELLO, M., DOAN-NGUYEN, V. V. T., GORDON, T. R., DIAZ, R. E., STACH, E. A., GORTE, R. J., FORNASIERO, P. & MURRAY, C. B. (2013). Control of Metal Nanocrystal Size Reveals Metal-Support Interface Role for Ceria Catalysts. *Science* **341**, 771–773.

CARRETTIN, S., CONCEPCIÓN, P., CORMA, A., NIETO, J. M. L. & PUNTES, V. F. (2004). Nanocrystalline CeO2 Increases the Activity of Au for CO Oxidation by Two Orders of Magnitude. *Angewandte Chemie International Edition* **43**, 2538–2540.

CHEE, S. W., ARCE-RAMOS, J. M., LI, W., GENEST, A. & MIRSAIDOV, U. (2020). Structural changes in noble metal nanoparticles during CO oxidation and their impact on catalyst activity. *Nature Communications* **11**.

CHENNA, S. & CROZIER, P. A. (2012). Operando Transmission Electron Microscopy: A Technique for Detection of Catalysis Using Electron Energy-Loss Spectroscopy in the Transmission Electron Microscope. *Acs Catalysis* **2**, 2395–2402.

COTTON, F. A. (1975). Fluxionality in organometallics and metal carbonyls. *Journal of Organometallic Chemistry* **100**, 29–41.

CROZIER, P. A. & CHENNA, S. (2011). In situ analysis of gas composition by electron energy-loss spectroscopy for environmental transmission electron microscopy. *Ultramicroscopy* **111**, 177–185.

CROZIER, P. A., LAWRENCE, E. L., VINCENT, J. L. & LEVIN, B. D. A. (2019). Dynamic Restructuring during Processing: Approaches to Higher Temporal Resolution. *Microscopy and Microanalysis* **25**, 1464–1465.

DATYE, A. K., KALAKKAD, D. S., YAO, M. H. & SMITH, D. J. (1995). Comparison of Metal-Support Interactions in Pt/TiO2 and Pt/CeO2. *Journal of Catalysis* **155**, 148–153.

GÄNZLER, A. M., CASAPU, M., MAURER, F., STÖRMER, H., GERTHSEN, D., FERRÉ, G., VERNOUX, P., BORNMANN, B., FRAHM, R., MURZIN, V., NACHTEGAAL, M., VOTSMEIER, M. & GRUNWALDT, J.-D. (2018). Tuning the Pt/CeO2 Interface by in Situ Variation of the Pt Particle Size. *ACS Catalysis* **8**, 4800–4811.

GÄNZLER, A. M., CASAPU, M., VERNOUX, P., LORIDANT, S., CADETE SANTOS AIRES, F. J., EPICIER, T., BETZ, B., HOYER, R. & GRUNWALDT, J. D. (2017). Tuning the Structure of Platinum Particles on Ceria In Situ for Enhancing the Catalytic Performance of Exhaust Gas Catalysts. *Angewandte Chemie - International Edition* **56**, 13078–13082.

GUERRERO-PÉREZ, M. O. & BAÑARES, M. A. (2006). From conventional in situ to operando studies in Raman spectroscopy. *Catalysis Today* **113**, 48–57.

HAN, J., CHEN, D. & ZHU, J. (2019). Isotopic Oxygen Exchange between CeO 2 and O 2 : A Heteroexchange Mechanism. *ChemistrySelect* **4**, 13280–13283.

HANSEN, P. L., WAGNER, J. B., HELVEG, S., ROSTRUP-NIELSEN, J. R., CLAUSEN, B. S. & TOPSØE, H. (2002). Atom-resolved imaging of dynamic shape changes in supported copper nanocrystals. *Science* **295**, 2053–2055.

HATANAKA, M., TAKAHASHI, N., TANABE, T., NAGAI, Y., DOHMAE, K., AOKI, Y., YOSHIDA, T. & SHINJOH, H. (2010). Ideal Pt loading for a Pt/CeO2-based catalyst stabilized by a Pt-O-Ce bond. *Applied Catalysis B: Environmental* **99**, 336–342.

JIA, C. L., HOUBEN, L., THUST, A. & BARTHEL, J. (2010). On the benefit of the negative-spherical-aberration imaging technique for quantitative HRTEM. *Ultramicroscopy* **110**, 500–505.

KAMIYA, M., SHIMADA, E., IKUMA, Y., KOMATSU, M. & HANEDA, H. (2000). Intrinsic and Extrinsic Oxygen Diffusion and Surface Exchange Reaction in Cerium Oxide. *Journal of The Electrochemical Society* **147**, 1222.

KOPELENT, R., VAN BOKHOVEN, J. A., SZLACHETKO, J., EDEBELI, J., PAUN, C., NACHTEGAAL, M. & SAFONOVA, O. V (2015). Catalytically Active and Spectator Ce3+ in Ceria-Supported Metal Catalysts. *Angewandte Chemie - International Edition* **54**, 8728–8731.

LANGDON, J. T., VINCENT, J. L. & CROZIER, P. A. (2019). Finite Element Modeling of Gas and Temperature Distributions during Catalytic Reactions in an Environmental Transmission Electron Microscope. *Microscopy and Microanalysis* **25**, 2014–2015.

LAWRENCE, ETHAN L, LEVIN, B. D. A., BOLAND, T., CHANG, S. L. Y. & CROZIER, P. A. (2021). Atomic Scale Characterization of Fluxional Cation Behavior on Nanoparticle Surfaces : Probing Oxygen Vacancy Creation / Annihilation at Surface Sites.

LAWRENCE, ETHAN L., LEVIN, B. D. A., BOLAND, T., CHANG, S. L. Y. & CROZIER, P. A. (2021). Atomic Scale Characterization of Fluxional Cation Behavior on Nanoparticle Surfaces: Probing Oxygen Vacancy Creation/Annihilation at Surface Sites. *ACS Nano*.

LEVIN, B. D. A., LAWRENCE, E. L. & CROZIER, P. A. (2020). Tracking the picoscale spatial motion of atomic columns during dynamic structural change. *Ultramicroscopy* **213**, 112978.

LI, P., CHEN, X., LI, Y. & SCHWANK, J. W. (2019). A review on oxygen storage capacity of CeO2-based materials: Influence factors, measurement techniques, and applications in reactions related to catalytic automotive emissions control. *Catalysis Today* **327**, 90–115.

LI, Y., KOTTWITZ, M., VINCENT, J. L., ENRIGHT, M. J., LIU, Z., ZHANG, L., HUANG, J., SENANAYAKE, S. D., YANG, W. D., CROZIER, P. A., NUZZO, R. G. & FRENKEL, A. I. (2021). Dynamic structure of active sites in ceria-supported Pt catalysts for the water gas shift reaction. *Nature Communications* **12**, 914.

LI, Y., ZAKHAROV, D., ZHAO, S., TAPPERO, R., JUNG, U., ELSEN, A., BAUMANN, P., NUZZO, R. G., STACH, E. A. & FRENKEL, A. I. (2015). Complex structural dynamics of nanocatalysts revealed in Operando conditions by correlated imaging and spectroscopy probes. *Nature Communications* **6**, 1–6.

LIU, H. H., WANG, Y., JIA, A. P., WANG, S. Y., LUO, M. F. & LU, J. Q. (2014). Oxygen vacancy promoted CO oxidation over Pt/CeO2 catalysts: A reaction at Pt-CeO2 interface. *Applied Surface Science* **314**, 725–734.

LU, Y., THOMPSON, C., KUNWAR, D., DATYE, A. K. & KARIM, A. M. (2020). Origin of the High CO Oxidation Activity on CeO2 Supported Pt Nanoparticles: Weaker Binding of CO or Facile Oxygen Transfer from the Support? *ChemCatChem* **12**, 1726–1733.

MAI, H. X., SUN, L. D., ZHANG, Y. W., SI, R., FENG, W., ZHANG, H. P., LIU, H. C. & YAN, C. H. (2005). Shape-selective synthesis and oxygen storage behavior of ceria nanopolyhedra, nanorods, and nanocubes. *Journal of Physical Chemistry B* **109**, 24380–24385.

MARS, P. & VAN KREVELEN, D. W. (1954). Oxidations carried out by means of vanadium oxide catalysts. *Chemical Engineering Science* **3**, 41–59.

MARTIN, D. & DUPREZ, D. (1996). Mobility of surface species on oxides. 1. isotopic exchange of 18O2 with 16O of SiO2, Al2O3, ZrO2, MgO, CeO2, and CeO2-Al2O3. activation by noble metals. correlation with oxide basicity. *Journal of Physical Chemistry* **100**, 9429–9438.

MCCRUM, I. T., HICKNER, M. A. & JANIK, M. J. (2017). First-Principles Calculation of Pt Surface Energies in an Electrochemical Environment: Thermodynamic Driving Forces for Surface Faceting and Nanoparticle Reconstruction. *Langmuir* **33**, 7043–7052.

MILLER, B. K., BARKER, T. M. & CROZIER, P. A. (2015). Novel sample preparation for operando TEM of catalysts. *Ultramicroscopy* **156**, 18–22.

MILLER, B. K. & CROZIER, P. A. (2014). Analysis of Catalytic Gas Products Using Electron Energy-Loss Spectroscopy and Residual Gas Analysis for Operando Transmission Electron Microscopy. *Microscopy and Microanalysis* **20**, 815–824.

——— (2021). Linking Changes in Reaction Kinetics and Atomic-Level Surface Structures on a Supported Ru Catalyst for CO Oxidation. *ACS Catalysis* 1456–1463.

MULLINS, D. R. (2015). The surface chemistry of cerium oxide. *Surface Science Reports* **70**, 42.

NAGAI, Y., HIRABAYASHI, T., DOHMAE, K., TAKAGI, N., MINAMI, T., SHINJOH, H. & MATSUMOTO, S. (2006). Sintering inhibition mechanism of platinum supported on ceria-based oxide and Pt-oxide-support interaction. *Journal of Catalysis* **242**, 103–109.

PEREIRA-HERNÁNDEZ, X. I., DELARIVA, A., MURAVEV, V., KUNWAR, D., XIONG, H., SUDDUTH, B., ENGELHARD, M., KOVARIK, L., HENSEN, E. J. M., WANG, Y. & DATYE, A. K. (2019). Tuning Pt-CeO2 interactions by high-temperature vapor-phase synthesis for improved reducibility of lattice oxygen. *Nature Communications* **10**.

PLODINEC, M., NERL, H. C., GIRGSDIES, F., SCHLÖGL, R. & LUNKENBEIN, T. (2020). Insights into Chemical Dynamics and Their Impact on the Reactivity of Pt Nanoparticles during CO Oxidation by Operando TEM. *ACS Catalysis* **10**, 3183–3193.

PODKOLZIN, S. G., SHEN, J., DE PABLO, J. J. & DUMESIC, J. A. (2000). Equilibrated Adsorption of CO on Silica-Supported Pt Catalysts. *Journal of Physical Chemistry B* **104**, 4169–4180.

PUIGDOLLERS, A. R., SCHLEXER, P., TOSONI, S. & PACCHIONI, G. (2017). Increasing oxide reducibility: the role of metal/oxide interfaces in the formation of oxygen vacancies. *ACS Catalysis* **7**, 6493–6513.

SALCEDO, A. & IRIGOYEN, B. (2020). Unraveling the Origin of Ceria Activity in Water-Gas Shift by First-Principles Microkinetic Modeling. *Journal of Physical Chemistry C* **124**, 7823–7834.

SHINJOH, H., HATANAKA, M., NAGAI, Y., TANABE, T., TAKAHASHI, N., YOSHIDA, T. & MIYAKE, Y. (2009). Suppression of noble metal sintering based on the support anchoring effect and its application in automotive three-way catalysis. *Topics in Catalysis* **52**, 1967–1971.

SINCLAIR, R., LEE, S. C., SHI, Y. & CHUEH, W. C. (2017). Structure and chemistry of epitaxial ceria thin films on yttria-stabilized zirconia substrates, studied by high resolution electron microscopy. *Ultramicroscopy* **176**, 200–211.

SOMORJAI, G. A. (1991). The Flexible Surface. Correlation between Reactivity and Restructuring Ability. *Langmuir* **7**, 3176–3182.

SUN, G. & SAUTET, P. (2018). Metastable Structures in Cluster Catalysis from First-Principles: Structural Ensemble in Reaction Conditions and Metastability Triggered Reactivity. *Journal of the American Chemical Society* **140**, 2812–2820.

TROVARELLI, A. (2002). *Catalysis by Ceria and Related Materials*. 2nd ed. Trovarelli, A. (Ed.). London: Imperial College Press.

VENDELBO, S. B., ELKJÆR, C. F., FALSIG, H., PUSPITASARI, I., DONA, P., MELE, L., MORANA, B., NELISSEN, B. J., VAN RIJN, R., CREEMER, J. F., KOOYMAN, P. J. & HELVEG, S. (2014). Visualization of oscillatory behaviour of Pt nanoparticles catalysing CO oxidation. *Nature Materials* **13**, 884–890.
VINCENT, J. & CROZIER, P. (2020). Atomic-resolution *Operando* and Time-resolved *In Situ* TEM Imaging of Oxygen Transfer Reactions Catalyzed by CeO2-supported Pt Nanoparticles. *Microscopy and Microanalysis* **26**, 1694–1695.
VINCENT, J. L., VANCE, J. W., LANGDON, J. T., MILLER, B. K. & CROZIER, P. A. (2020). Chemical Kinetics for Operando Electron Microscopy of Catalysts: 3D Modeling of Gas and Temperature Distributions During Catalytic Reactions. *Ultramicroscopy* 113080.
WINTERBOTTOM, W. L. (1967). Equilibrium shape of a small particle in contact with a foreign substrate. *Acta Metallurgica* **15**, 303–310.
WU, Z., LI, M., HOWE, J., MEYER, H. M. & OVERBURY, S. H. (2010). Probing defect sites on CeO2 nanocrystals with well-defined surface planes by raman spectroscopy and O2 adsorption. *Langmuir* **26**, 16595–16606.
WU, Z., LI, M. & OVERBURY, S. H. (2012). On the structure dependence of CO oxidation over CeO2 nanocrystals with well-defined surface planes. *Journal of Catalysis* **285**, 61–73.
ZHAI, H. & ALEXANDROVA, A. N. (2017). Fluxionality of Catalytic Clusters: When It Matters and How to Address It. *ACS Catalysis* **7**, 1905–1911.
ZHANG, Z., ZANDKARIMI, B. & ALEXANDROVA, A. N. (2020). Ensembles of Metastable States Govern Heterogeneous Catalysis on Dynamic Interfaces. *Accounts of Chemical Research* **53**, 447–458.
ZHU, B., MENG, J., YUAN, W., ZHANG, X., YANG, H., WANG, Y. & GAO, Y. (2020). Reshaping of Metal Nanoparticles Under Reaction Conditions. *Angewandte Chemie - International Edition* **59**, 2171–2180.

**Figures and captions**

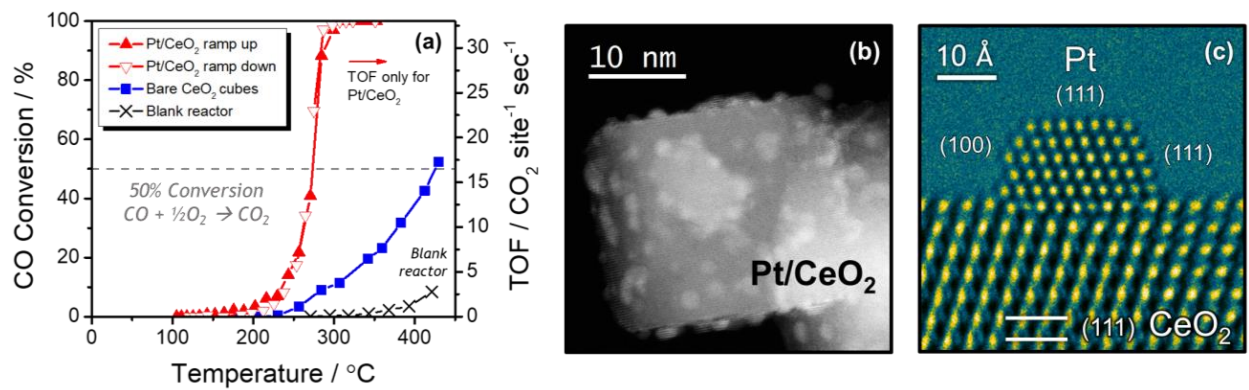

**Figure 1. (a)** Catalytic activity for CO oxidation evaluated in a packed-bed plug-flow reactor. The activity of the bare $CeO_2$ support is demonstrably less and the effect of blank reactor is negligible. **(b)** Z-contrast STEM image of a typical Pt-loaded $CeO_2$ nanoparticle showing dispersion of roughly 1.6 nm Pt nanoparticles loaded on the $CeO_2$ nanoparticle support. **(c)** HRTEM image of a typical Pt nanoparticle supported on a (111) surface of a $CeO_2$ nanoparticle, with indices of exposed Pt surfaces marked in the image.

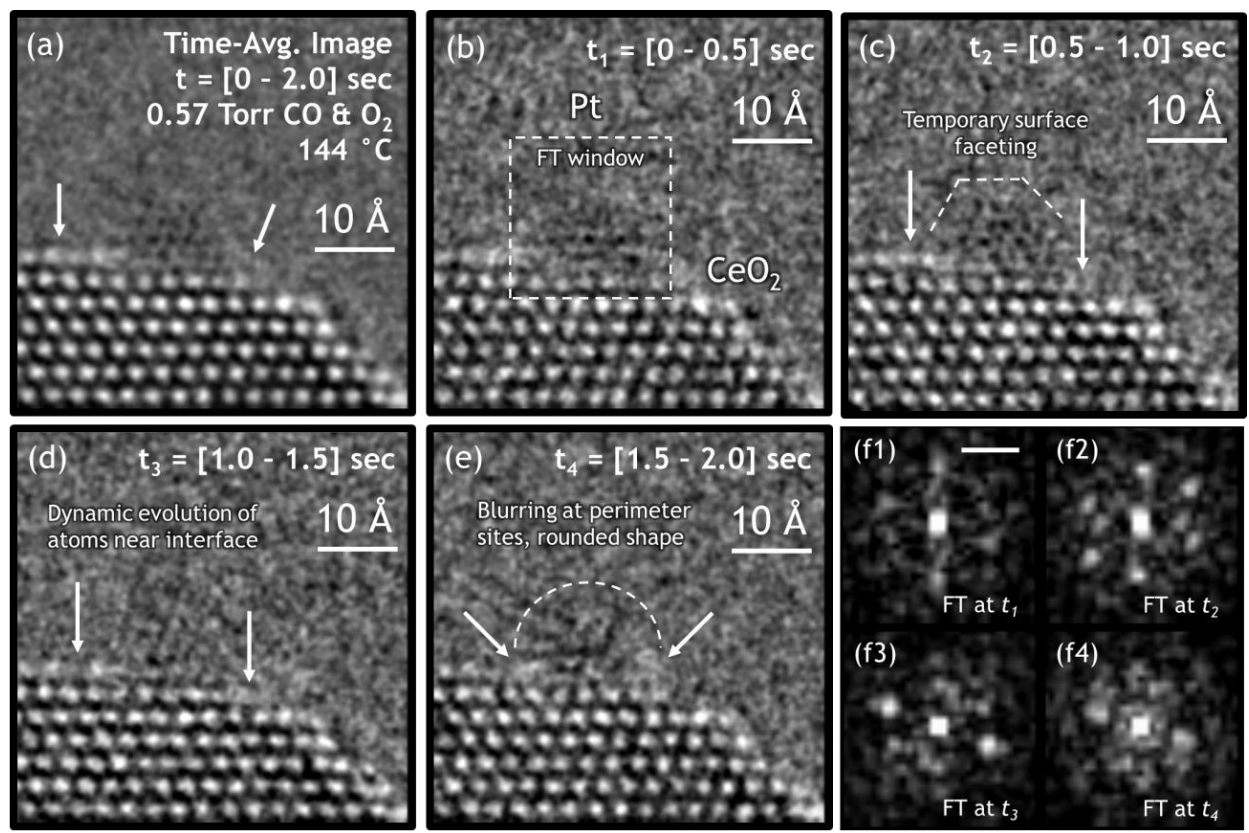

**Figure 2.** *In situ* ETEM image time-series of $CeO_2$-supported Pt NP at 144 °C in 0.57 Torr of CO and $O_2$. Part **(a)** shows the time-averaged image of the catalyst, obtained by summing together the individual 0.5 second exposure frames over the entire [0 – 2.0] second acquisition period. Parts **(b)** – **(e)** show the atomic-scale structural dynamics that evolve over 0.5 second intervals from $t = 0$ seconds to $t = 2.0$ seconds. Parts **(f1)** – **(f4)** display the FT taken at each time interval from the windowed region around the Pt NP, as denoted in **(b)**. The scale bar in **(f1)** is 5.0 nm$^{-1}$. Images have been processed with a bandpass filter for clarity. FTs were produced from unfiltered, windowed images that were processed with a Hanning function to remove edge artifacts caused by windowing; the modulus of the FT is shown.

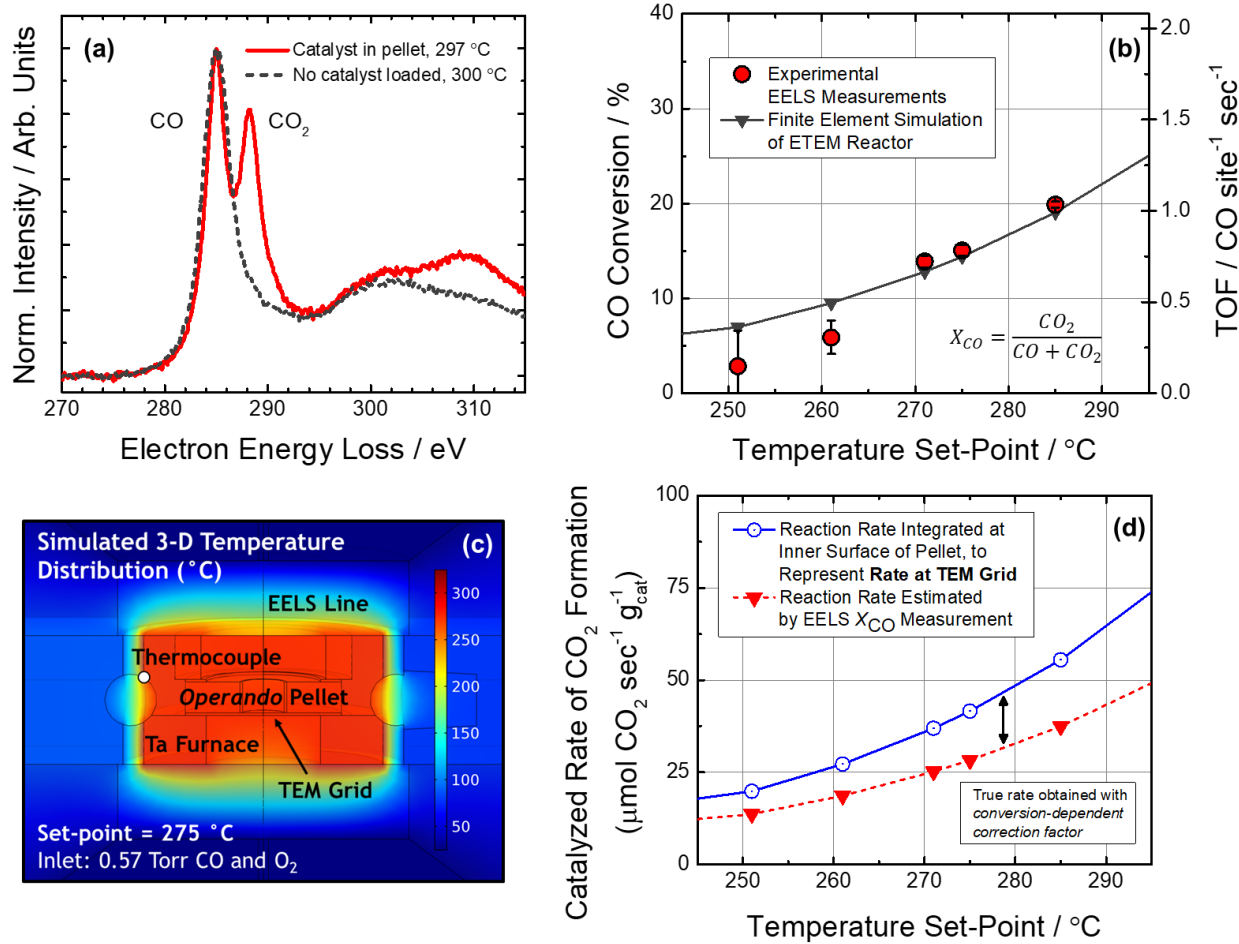

**Figure 3.** Detection and quantification of *in situ* CO conversion and catalytic reaction rate in the ETEM reactor. In **(a)** a set of background-subtracted electron energy-loss spectra taken from the gas atmosphere around the reactor demonstrates that the conversion of CO (C π* peak at 285 eV) into $CO_2$ (C π* peak at 288.3 eV) is attributed unambiguously to the $Pt/CeO_2$ catalyst, not the reactor. In **(b)** the CO conversion detected with EELS during the *operando* ETEM experiment was quantified and plotted as a function of temperature (red circles). The solid gray curve displays the CO conversion evaluated under nominally identical conditions within a finite element simulation of the ETEM reactor. In **(c)** the simulated temperature distribution in the model reactor is plotted for a furnace set point of 275 °C. Finally, in **(d)** an analysis of the catalyzed rate of product formation in the model is presented, which provides a conversion-dependent correction factor for quantitatively determining the catalytic reaction rate from the *in situ* conversion measurement.

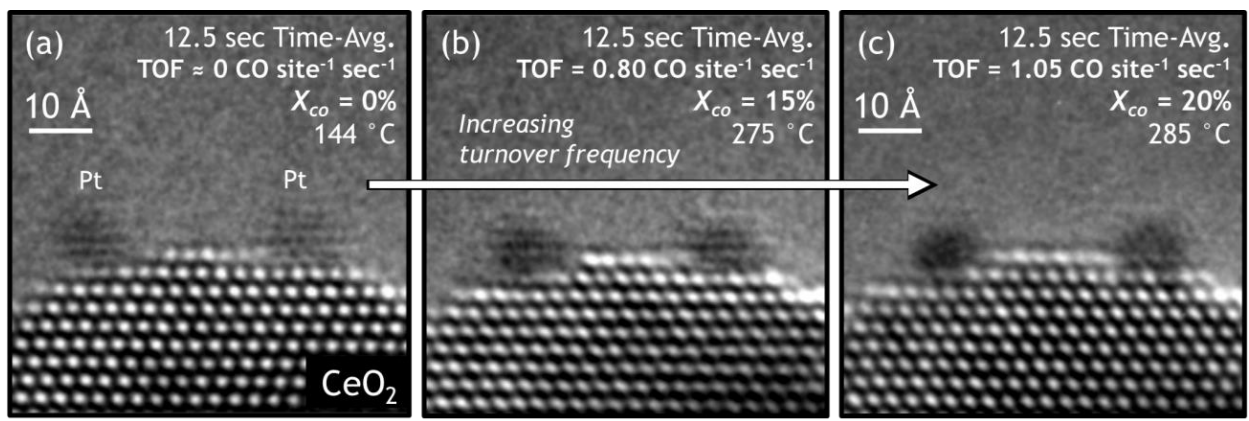

**Figure 4.** 12.5 second time-averaged *operando* TEM images showing the dynamic structural evolution of the Pt/CeO$_2$ interface and nearby support surface for CO conversions of **(a)** $X_{co} = 0\%$, **(b)** $X_{co} = 15\%$, and **(c)** $X_{co} = 20\%$. The corresponding temperatures and perimeter-site normalized TOFs are stated in the respective figures.

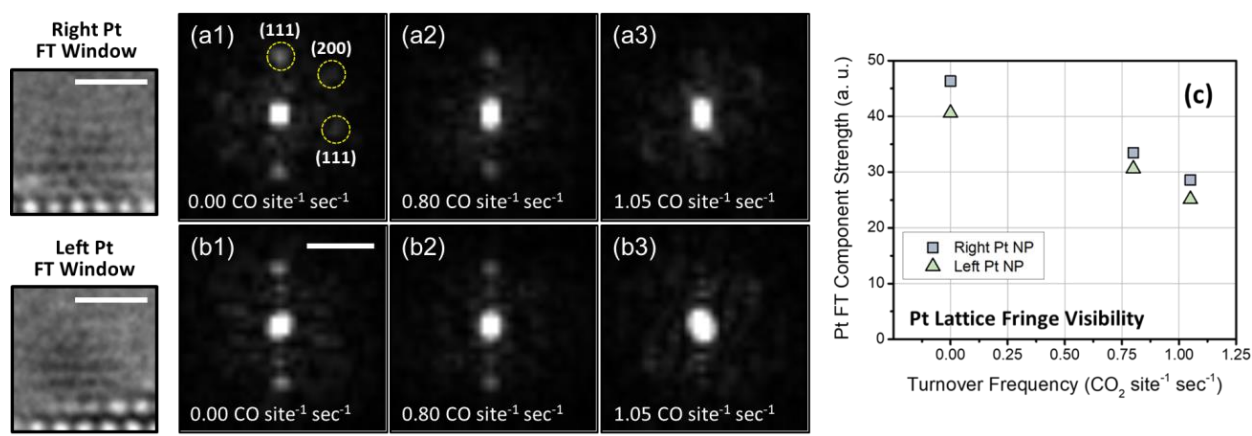

**Figure 5.** Quantification of Pt nanoparticle (NP) fringe visibility and correlation with turnover frequency for CO oxidation. The fringe visibility is determined through FT analysis of windowed regions centered on the Pt NPs shown, e.g., at left. Scale bars in images correspond to 1 nm. Parts **(a1)** – **(a3)** show the modulus of the FTs after applying a Hanning filter to the windowed images of the right Pt NP at conditions corresponding to TOFs of 0.00, 0.80, and 1.05 CO site$^{-1}$ sec$^{-1}$, respectively. Parts **(b1)** – **(b3)** show the modulus of the FTs taken from the left Pt NP at the same respective conditions. The scale bar in **(b1)** corresponds to 5.0 nm$^{-1}$. In **(c)**, the Pt fringe visibility is quantified and plotted as a function of catalytic turnover frequency.

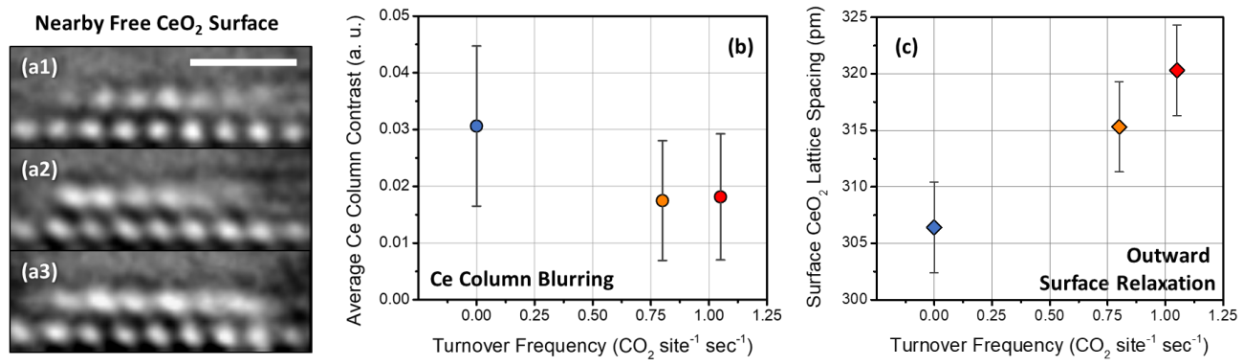

**Figure 6.** Quantification of fluxional behavior taking place on the free $CeO_2$ (111) surface neighboring the Pt nanoparticles and correlation with TOF for CO oxidation. Enhanced views of the nearby free $CeO_2$ (111) surface are shown for TOFs of **(a1)** 0.00 CO site$^{-1}$ sec$^{-1}$, **(a2)** 0.80 CO site$^{-1}$ sec$^{-1}$, and **(a3)** 1.05 CO site$^{-1}$ sec$^{-1}$. The scale bar in **(a1)** corresponds to 1 nm. Part **(b)** displays the average surface Ce atomic column contrast plotted against catalytic TOF; error bars are given as the standard deviation relative to the mean measurement. In **(c)**, the separation distance between the surface and subsurface $CeO_2$ (111) lattice planes is correlated with catalytic TOF, revealing an outward surface relaxation that grows larger with increasing activity. Error bars are derived from the standard deviation about the mean lattice spacing measured in the subsurface and bulk $CeO_2$.

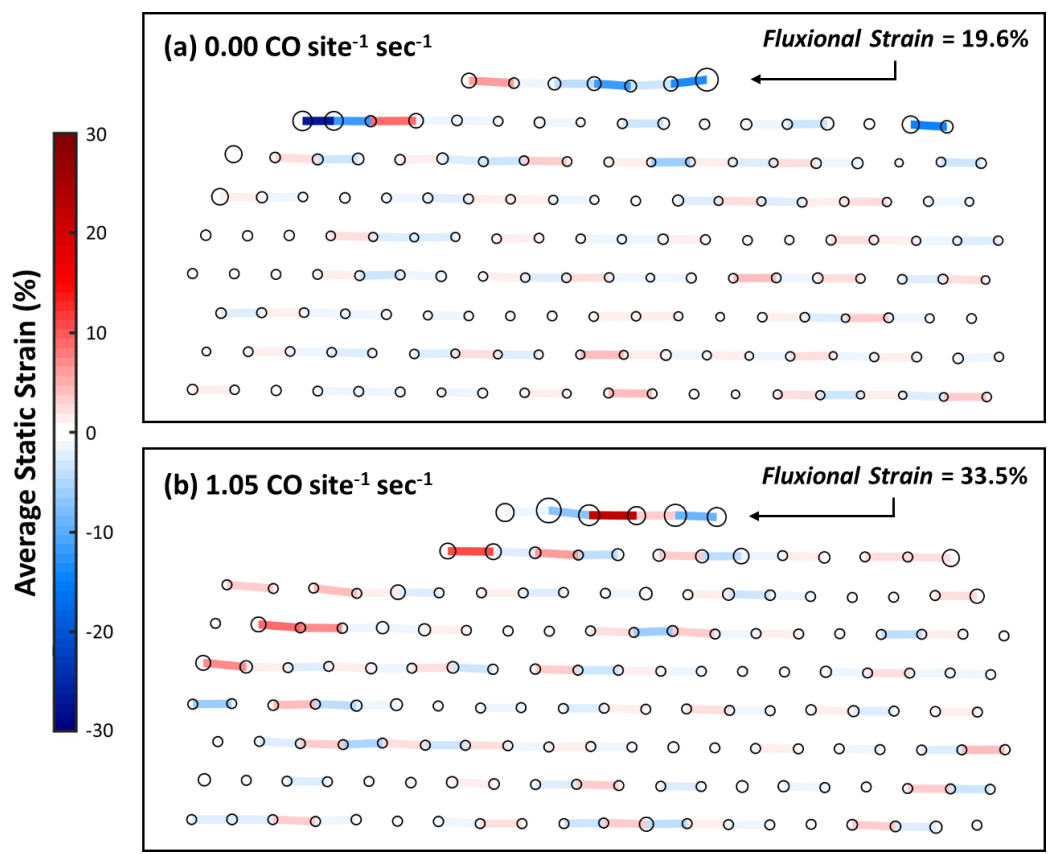

**Figure 7.** Average static and fluxional strain in $CeO_2$ support at TOFs of **(a)** 0.00 CO site$^{-1}$ sec$^{-1}$ and **(b)** 1.05 CO site$^{-1}$ sec$^{-1}$. The circles signify the Ce atomic column position determined by Gaussian peak fitting and are drawn proportional in size to the standard deviation of the fitted Gaussian. The strain between sites is represented with a colored bar; red and blue correspond to tensile and compressive strain, respectively. The free $CeO_2$ (111) surface terrace between the supported Pt nanoparticles has been marked with a black arrow.

# Supplementary Information

Fluxional Behavior at the Atomic Level and its Impact on Activity: CO Oxidation over $CeO_2$-supported Pt Catalysts

Joshua L. Vincent[1], and Peter A. Crozier[1]*

[1]*School for Engineering of Matter, Transport, and Energy, Arizona State University, Tempe, Arizona 85281*

*Corresponding Author:
Peter A. Crozier
Engineering G Wing, #301
501 E. Tyler Mall
Tempe, AZ 85287-6106
Tel: 480 965 2934
Fax: 480 727 9321
Email: crozier@asu.edu

# 1. Catalyst preparation and characterization

The catalyst consists of Pt nanoparticles supported on $CeO_2$ nanoparticles having a generally cubic morphology. Nanostructured $CeO_2$ cubes were chosen as model support due to their shape which facilitates the identification of on-axis particles during high resolution imaging. Methods describing the synthesis procedures for various $CeO_2$ morphologies have been published previously (Mai et al., 2005). Generally, the synthesis involves the precipitation of crystalline $CeO_2$ from an aqueous basic solution when heated autogenously in a hydrothermal vessel. In a typical synthesis, 1.38 g of $Ce(NO_3)_3 \cdot 6H_2O$ dissolved in 8 mL of deionized (DI) water was added to a 60 mL solution of 12 M NaOH, which was stirred for 30 minutes. The resultant pale-white slurry was added to an 85 mL Teflon-lined autoclave and heated for 24 hours at 200 °C. Upon cooling to room temperature, the fluffy white precipitates were isolated by filtration, washed multiple times with DI water, and dried at 60 °C in air overnight. After drying, the powders were calcined in air at 350 °C for 4 hours.

An impregnation technique was used to deposit 17 wt.% Pt onto the $CeO_2$ nanocubes. A high weight loading of metal was desired to reduce the amount of time spent searching for Pt nanoparticles close to a zone-axis orientation during *in situ* and *operando* TEM experiments. In brief, a volume of $CeO_2$ powder was weighed, then an appropriate mass of $H_2PtCl_6$ was dissolved in DI water to achieve the desired weight loading. This Pt-containing solution was added dropwise to the $CeO_2$ to form a slurry. The slurry was mixed continuously in a mortar and pestle for 2 hours until the complete evaporation of the DI water was achieved. The powder was dried overnight in air at 60 °C and then reduced in a flowing stream of 5% $H_2$/Ar for 2 hours at 400 °C, yielding a black powdered sample of $Pt/CeO_2$. Bulk structural characterization was performed using X-ray diffraction (XRD) on a Bruker D-5000 with a Cu K$\alpha$ source ($\lambda = 0.15406$ nm). The powder XRD patterns of the bare $CeO_2$ and Pt-loaded $CeO_2$ nanocubes are shown in **Figure S1**. A simulated XRD pattern of an infinite crystal of $CeO_2$ is also shown for reference (*JCPDS* No. 34-0394). The peaks present in the XRD pattern for the bare $CeO_2$ nanocubes (blue line) match well with that of the simulated crystal (black line), indicating that the sample is phase-pure $CeO_2$ (space group Fm-3m, a = 5.41 Å). The XRD pattern for the $Pt/CeO_2$ nanocubes (red line) is essentially identical to that of the bare $CeO_2$ support.

The as-reduced $Pt/CeO_2$ catalyst powder was imaged in a probe-corrected JOEL ARM 200F scanning transmission electron microscope (STEM) operated at 200 kV. A TEM specimen was prepared by dry-loading the freshly reduced catalyst powder onto a holey carbon film. High angle annular dark field (HAADF) or so-called Z-contrast images were collected of many different areas of the TEM specimen and the Pt nanoparticle size distribution was determined by measuring the size of about 500 different Pt nanoparticles. **Figure S2** displays two representative HAADF STEM images of the $Pt/CeO_2$ powder showing the dispersion of the Pt on the $CeO_2$ support along with a histogram of the size distribution measurement. The histogram of the size distribution measurement shows that the Pt nanoparticle population has an average nanoparticle size of 1.6 nm.

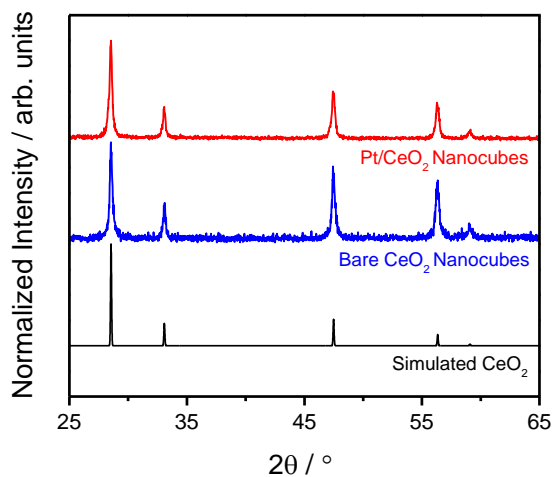

**Figure S1.** Powder XRD patterns from (blue) bare CeO$_2$ and (red) Pt-loaded CeO$_2$ compared with (black) a simulated pattern from a perfect CeO$_2$ crystal.

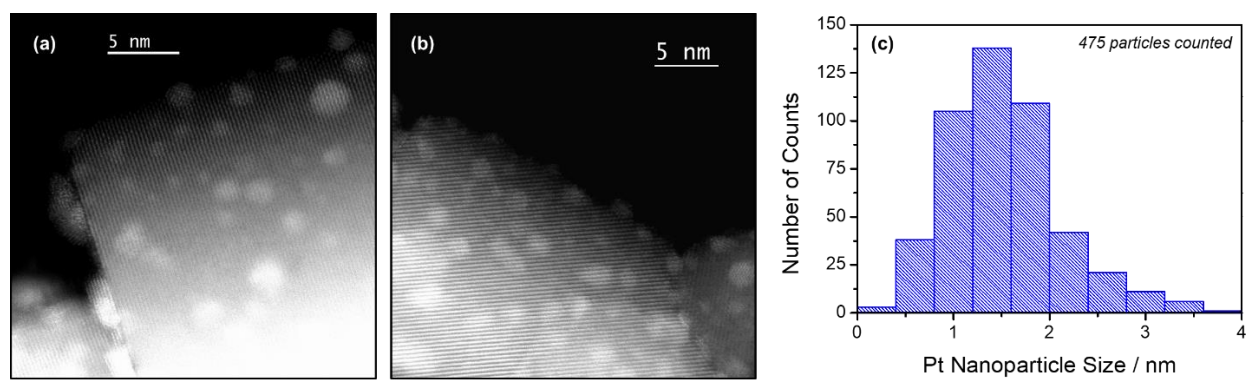

**Figure S2. (a, b)** Z-contrast STEM images of a typical Pt-loaded CeO$_2$ nanoparticle imaged during the Pt nanoparticle size measurement, and **(c)** Pt nanoparticle size distribution histogram (n = 475) for the 17 wt.% Pt/CeO$_2$ catalyst, which exhibits an average Pt particle size of 1.6 nm.

The catalyst's activity for CO oxidation was evaluated in a packed bed plug flow reactor. Plug flow reactor experiments were performed in a RIG-150 micro-reactor from In Situ Research Instruments (ISRI). Effluent gas compositions were measured with a Varian 3900 gas chromatograph (GC) equipped with a thermal conductivity detector (TCD). In a typical reaction, 20 mg of catalyst powder was diluted on 2.6 g of inert SiO$_2$ sand particles having size of 1 mm and loaded into the reactor. A reducing treatment was preformed prior to the introduction of reactants, wherein 40 standard cubic centimeters per minute (SCCM) of 5% H$_2$/Ar was flown while the reactor bed was held at 400 °C for 2 hours. The reactor was then cooled to room temperature with continued 5% H$_2$/Ar flow. Upon cooling, the gas stream was switched to 8 SCCM of 10% CO/He and 22 SCCM of 5% O$_2$/He, balanced with 150 SCCM of pure He, which corresponded to a gas hourly space velocity of 12,500 hr$^{-1}$.

The reactor was then heated up to 400 °C at a rate of 1.5 °C/minute while the effluent gas composition was analyzed. He gas was used as a carrier gas. The reaction temperature was monitored by a thermocouple placed in the center of the catalyst bed. The CO conversion, $X_{CO}$, was quantified by calculating $X_{CO} = (CO_{in}-CO_{out})/CO_{in}$, where $CO_{in}$ and $CO_{out}$ denote the molar flowrate of CO into and out of the reactor, respectively. A reaction rate was calculated by multiplying the conversion by the molar flow rate of CO into the reactor. Turnover frequencies were computed by normalizing the reaction rates to the estimated number of Pt atoms at the metal-support interfacial perimeter, as determined by the derivation described in the next section of the Supplemental Information. We normalize to the number of Pt atoms at the perimeter as we assume that the reaction occurs through a Mars-van Krevelen mechanism at the perimeter of the metal-support interface. We note that this normalization assumes that each perimeter Pt atom is active and that the number of perimeter Pt atoms is identical both in vacuum conditions, where the number is counted, and under reaction conditions, where the catalysis takes place. Essentially, though, the TOF reported here is linearly related to the mass-normalized rate of product formation. Activation energies for CO oxidation were calculated through an Arrhenius analysis of the activity data taken in the low-conversion regime (i.e., < 25%). Figure 1a of the main text shows the light-off curves for CO oxidation. **Supplemental Figure S3** below shows the Arrhenius analysis of the rate data, which reveals that the Pt/CeO$_2$ catalyst shows an apparent activation energy $E_a$ of 74 kJ/mol.

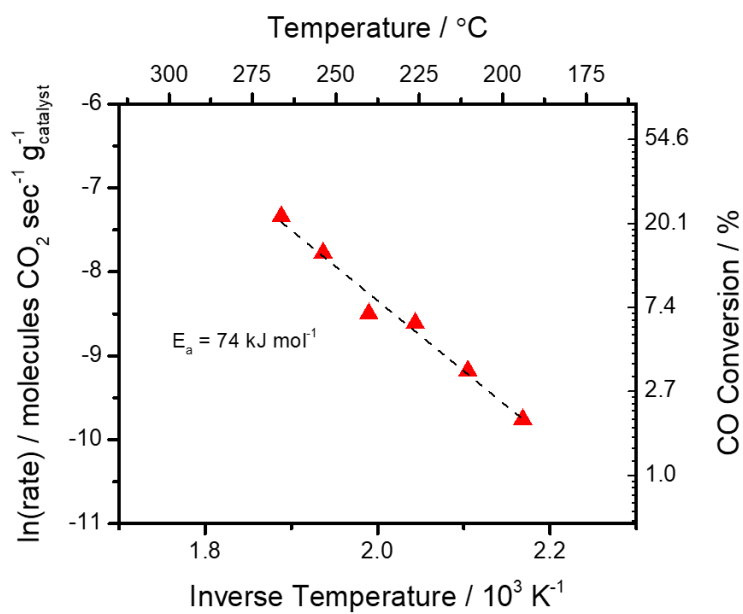

**Figure S3.** Arrhenius analysis of the Pt/CeO$_2$ catalyst light-off conversion data (see Figure 1a of the main text) shows that the apparent $E_a$ for CO oxidation is 74 kJ mol$^{-1}$.

## 2. Derivation of turnover frequency (TOF) on an interfacial perimeter-site basis from Pt nanoparticle size distribution

The number of atoms of Pt at the metal-support interfacial perimeter can be determined from the particle size histogram by assuming that the Pt nanoparticles have a hemispherical shape (Ertl et al., 1997). A derivation of this quantity and a sample calculation of the TOF for a given conversion are provided below. First, we define the TOF as:

(S1) $\quad TOF \left[\frac{molecules\ CO_2}{perimeter\ site * second}\right] = \frac{F_{CO,in} \times X_{CO}}{m_{Pt/CeO_2} \times \gamma}$

Where $F_{CO,in}$ represents the molar flowrate of CO into the *ex situ* reactor or the environmental cell, $X_{CO}$ represents the CO conversion, $m_{Pt/CeO2}$ represents the mass of Pt/CeO$_2$ catalyst loaded into the *ex situ* reactor or *operando* pellet, and $\gamma$ represents the number of atoms of Pt at the perimeter of the metal-support interface *per gram of Pt/CeO$_2$ catalyst*. The quantities $F_{CO,in}$ and $m_{Pt/CeO2}$ are controllable, and $X_{CO}$ is determined experimentally. To calculate the TOF for a given $X_{CO}$, one simply requires an estimate of the number of perimeter sites per gram of catalyst, i.e., $\gamma$.

The value of $\gamma$ was determined from HAADF-STEM measurements by assuming that the Pt nanoparticles have a hemispherical shape (Ertl et al., 1997). First, the volume, $V_i$, and interfacial perimeter length (i.e., circumference), $L_i$, was calculated for each particle $i$ measured with diameter $d_i$, as described in the following equations:

(S2) $\quad V_i = \frac{2}{3}\pi \left(\frac{d_i}{2}\right)^3$ (S3) $\quad L_i = \pi d_i$

The total volume, $V_{total}$, and interfacial perimeter length, $L_{total}$, of all the particles measured in the data set is then computed as the sum of all $n$ measurements (here $n = 475$):

(S4) $\quad V_{total} = \sum_{i=1}^{n} V_i$ (S5) $\quad L_{total} = \sum_{i=1}^{n} L_i$

Using the density of Pt, $\rho_{Pt} = 21.45$ g/cm$^3$, the total mass of Pt observed in the measurements is:

(S6) $\quad m_{total} = V_{total} \times \rho_{Pt}$

This allows for a mass-specific interfacial perimeter length, $L'_{total}$, to be calculated as:

(S7) $\quad L'_{total} = L_{total} \div m_{total} \times 0.17 \left[\frac{g_{Pt}}{g_{Pt/CeO_2}}\right]$

The factor of 0.17 is included since the catalyst is 17% Pt by weight. The last quantity needed to calculate the number of perimeter Pt atoms per gram catalyst is a value for the width of a Pt atom, $d_{Pt\ atom}$. Given the mass density of Pt, $\rho_{Pt}$, we can estimate a value by first calculating the number of atoms in a cubic nm of Pt, *i.e.*, the atomic volume density, $\varphi_{Pt}$:

(S8) $\quad \varphi_{Pt} = \rho_{Pt} \div M_{Pt} \times N_{Avogadro}$

Where $M_{Pt}$ is the molar mass of Pt, 195.08 g mol$^{-1}$, and $N_{Avogadro}$ is Avogadro's number, $6.02 \times 10^{23}$ atoms mol$^{-1}$. The width of a Pt atom can be computed by inverting and taking the cube root of the atomic volume density:

(S9) $\quad d_{Pt\ atom} = \left(\dfrac{1}{\varphi_{Pt}}\right)^{\frac{1}{3}}$

And so, the number of perimeter Pt atoms per gram catalyst, $\gamma$, is:

(S10) $\quad \gamma = L'_{total} \div d_{Pt\ atom}$

Following this arithmetic and performing the analysis on the Pt nanoparticle size distribution plotted in **Figure S2c**, a value for $\gamma$ of $5.45 \times 10^{19}$ sites g$^{-1}$ is obtained. A TOF can now be calculated for a given conversion assuming the CO flowrate and mass of catalyst in the reactor are known. During the *operando* TEM experiment, around 180 ug of Pt/CeO$_2$ catalyst was loaded into the operando pellet. The inlet flow rate of CO into the cell was 0.08 SCCM or $5.94 \times 10^{-8}$ mol CO sec$^{-1}$, which is $3.58 \times 10^{16}$ molecules CO sec$^{-1}$. It should be briefly mentioned here that the calculated TOF will still need to be corrected by finite element simulations in order to accurately represent the reaction rate of catalyst particles supported on the TEM grid, which is discussed at length in the following section of the SI. Regardless of this additional complexity, for an example CO conversion of 10%, the *in situ* TOF would be calculated as:

(S11) $\quad TOF = \dfrac{F_{CO,in} \times X_{CO}}{m_{Pt/CeO2} \times \gamma} = \dfrac{3.58 * 10^{16}\ [\text{molecules CO sec}^{-1}] \times 0.10}{180 * 10^{-6}\ [g] \times 5.45 \times 10^{19}\ \left[\frac{sites}{g}\right]} \approx 0.37 \left[\dfrac{molecules\ CO_2}{perimeter\ site * second}\right]$

The TOFs reported in this document have all been calculated following this procedure.

## 3. Details of finite element simulation of *operando* ETEM reactor

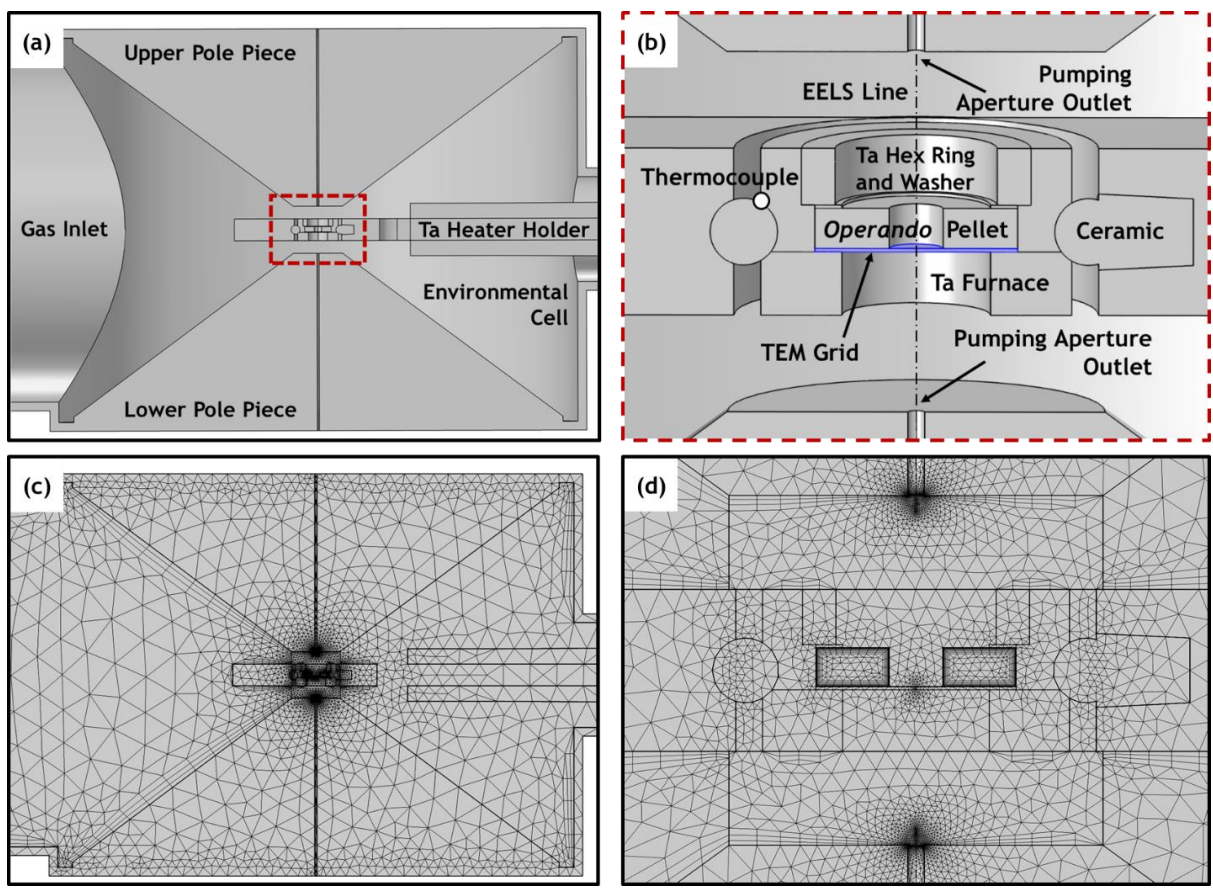

**Figure S4.** Finite element model geometry and mesh in full view of the environmental cell **(a, c)** as well as in an enhanced view focused on the *operando* ETEM reactor **(b, d)**. In **(b)**, the TEM grid domain is colored blue for clarity. A defined pressure and composition of reactant gas flows into the cell from the inlet on the left. The differential pumping aperture outlets in the pole pieces serve as outlets. **(b)** The pellet reactor and furnace holder reside between the two pole pieces. A thermocouple on the outer edge of the Ta furnace is used to control the temperature of the reactor.

Finite element simulations of the *operando* ETEM reactor were performed in COMSOL Multiphysics® software using an adapted model developed previously by Vincent et al. (Vincent et al., 2020). The Computational Fluid Dynamics, Heat Transfer, and Chemical Reaction Engineering modules were used. The model geometry is shown in the top of **Figure S4**, which displays (a) the environmental cell and (b) the *operando* ETEM pellet reactor. The TEM grid is highlighted in blue. The finite element mesh for the model is displayed in the bottom half of **Figure S4**, which shows the mesh, for (c) the full cell and (d) the reactor. The mesh is made of around 205,000 elements total.

Steady-state simulations were performed under conditions nominally identical to the *operando* ETEM experiment described in the main text. A reactant gas inflow of CO and $O_2$ in a 1:1.375 ratio was admitted to the cell. The total inlet flowrate was adjusted to 0.08 standard cubic centimeters per minute (SCCM) in order achieve a cell static pressure of 0.57 Torr. Heat transfer and multicomponent mass transport were implemented as described in (Vincent et al., 2020). The reaction was modeled as $0^{th}$ order, with an activation energy, $E_a$, of 74 kJ mol$^{-1}$, which was taken from the Arrhenius analysis done on the plug flow reactor data as presented in **Supplemental Figure S3**. The spatial distribution of catalyst in the pellet was modeled with an egg-shell profile. The furnace temperature was set in the range of 144 – 297 °C.

In the model, any quantity of interest (e.g., gas composition, temperature, reaction rate, etc.) may be determined at any element or averaged/integrated over chosen domains. For example, the rate of product formation (i.e., the reaction rate, with units of mol $CO_2$ per second) may be found by integrating the reaction rate throughout the domain of the *operando* pellet where the reaction occurs. Moreover, by integrating the rate along the inner surface of the pellet, where the composition and temperature are both nearly identical to that at the TEM grid, one can determine the reaction rate of the catalytic particles on the TEM grid (Vincent et al., 2020). As another example, the gas composition measured experimentally can be replicated as an integral of the composition along the path labeled "EELS Line" which represents the fast electron beam trajectory in the ETEM (see **Figure S4b**). A line integral along the EELS line can be used to simulate the reactant conversion that one would measure experimentally with EELS. Here, we have employed the model to establish a framework that allows us to link the reactant conversion measured along the EELS line to the reaction rate of the catalyst that is imaged on the TEM grid.

In the model, the true rate of product formation may be found by integrating the reaction rate throughout the domain of the pellet. The rate may be normalized to mass by also integrating the mass distribution within the same domain. Here we set the total integrated mass of the catalyst in the pellet to be 180 μg, as this corresponds to the amount loaded experimentally. As mentioned, one can determine the mass-normalized rate for the catalytic nanoparticles on the TEM grid by integrating the mass and rate along the innermost surface of the pellet (Vincent et al., 2020). Here we integrate the mass and rate around a 50 μm thick layer at the surface of the inner hole in the pellet. This rate, which refer to as $r_{grid}$, is tabulated as function of temperature in **Table S1**.

Experimentally, the rate of product formation may be estimated from the EELS CO conversion measurement. The estimated rate of $CO_2$ formation, $r_{EELS}$, may be calculated by multiplying the measured EELS CO conversion with the inlet molar flow rate of CO into the cell:

(S1) $\quad r_{EELS} = \frac{X_{CO} \times \dot{n}_{CO,in}}{m_{cat}}$

Here, $r_{EELS}$ is the estimated rate of $CO_2$ formation, $X_{CO}$ is the CO conversion measurement derived from EELS, $\dot{n}_{CO,in}$ is the inlet molar flow rate of CO into the cell, and $m_{cat}$ is the mass of catalyst loaded in the reactor. The mass-normalized rate of $CO_2$ formation estimated by the EELS CO conversion measurement may also be simulated in the model, since it is possible to replicate the conversion measurement by integrating the composition along the EELS line. The mass-normalized rate estimated by the EELS conversion measurement in the model is tabulated as a function of temperature in **Table S1**. The CO conversion values estimated through the EELS composition measurements are also given for reference.

It is of interest to compare the estimated – and experimentally measurable – rate of product formation, $r_{EELS}$, to the value which represents the rate at the TEM grid, $r_{grid}$. The last column of **Table S1** presents the calculated ratio of the two values. Observe that the ratio of $r_{EELS}$ to $r_{grid}$ is not constant with conversion but varies by nearly 20% from 0.80 to 0.66 over the conditions explored here. Previously we have shown that under higher conversion conditions (e.g., $X_{CO} >$ 0.70), the difference between the $r_{EELS}$ and $r_{grid}$ can grow beyond 200%, clearly demonstrating the need for and power of a model that allows one to relate the reactant conversion to the true reaction rate (i.e., activity) of the imaged catalyst. In the context of the present work, the true rate can be calculated correctly simply by scaling the rate measured through EELS by a conversion-dependent factor that is given in in the last column of **Table S1**.

**Table S1.** Summary of reaction rate analysis. The CO conversion estimated through the EELS measurement is presented as a function of the furnace temperature set point. The mass-normalized rate of $CO_2$ formation obtained by integrating the reaction rate and mass within a 50 μm thick layer at the surface along the inner hole in the pellet, $r_{grid}$, which represents the rate at the TEM grid, is presented, along with the rate estimated from the EELS CO conversion measurement, $r_{EELS}$. Finally, the ratio of the rate estimated through EELS and the rate at the TEM grid is given.

| Furnace Set Point (°C) | $X_{CO}$ Estimated from EELS (%) | $r_{grid}$ (μmol $CO_2$ sec$^{-1}$ g$_{cat}$$^{-1}$) | $r_{EELS}$ (μmol $CO_2$ sec$^{-1}$ g$_{cat}$$^{-1}$) | Ratio of $r_{EELS} : r_{grid}$ (%) |
|---|---|---|---|---|
| 144 | 0.11 | 0.26 | 0.21 | 80% |
| 202 | 1.26 | 3.48 | 2.44 | 70% |
| 251 | 7.07 | 19.83 | 13.66 | 69% |
| 261 | 9.65 | 27.20 | 18.65 | 69% |
| 271 | 13.01 | 36.89 | 25.16 | 68% |
| 275 | 14.61 | 41.55 | 28.25 | 68% |
| 285 | 19.33 | 55.51 | 37.40 | 67% |
| 297 | 26.61 | 77.57 | 51.53 | 66% |

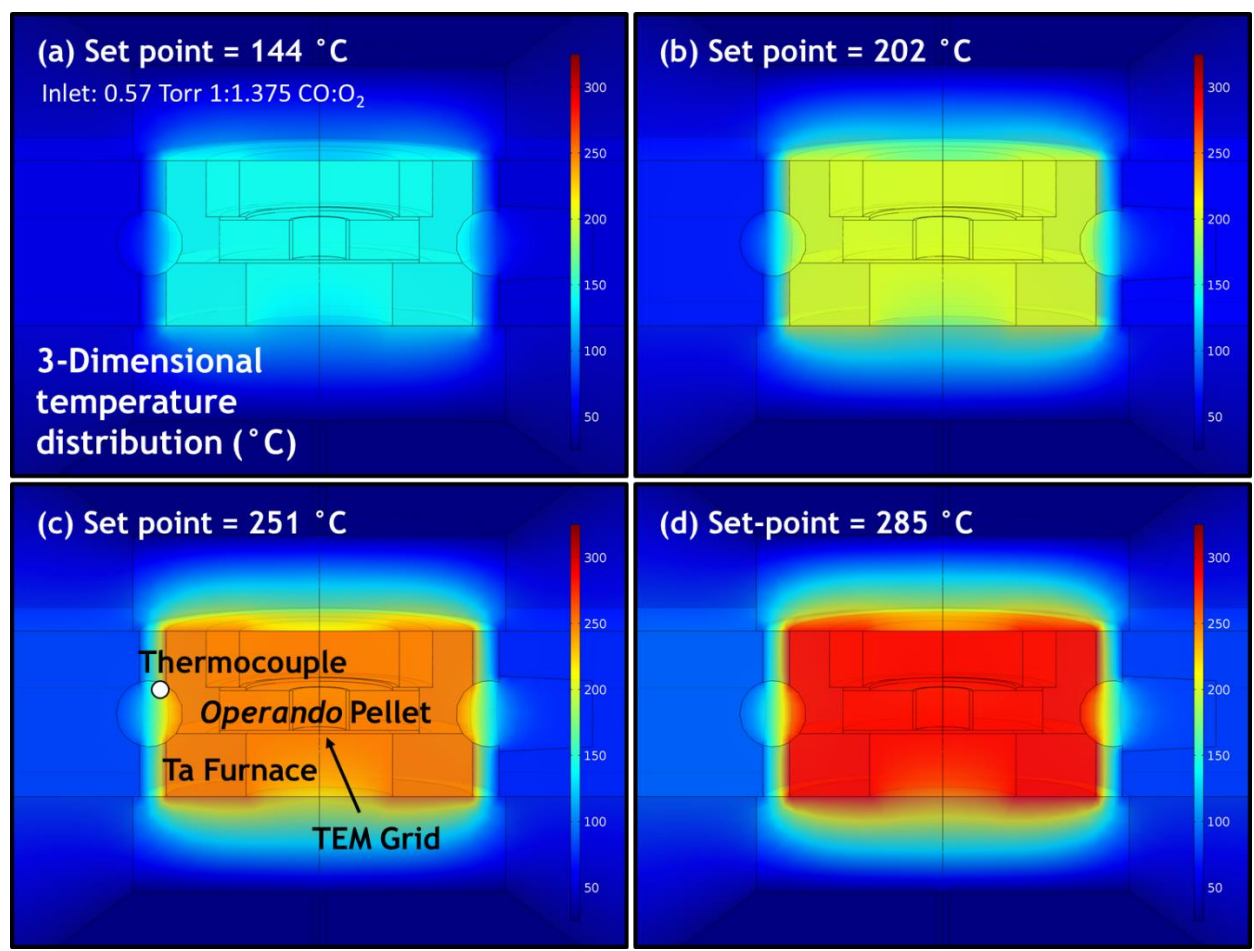

**Figure S5.** Temperature distribution in and around the ETEM reactor during catalysis for furnace thermocouple set points of **(a)** 144 °C, **(b)** 202 °C, **(c)** 251 °C, and **(d)** 285 °C. At each condition the temperature distribution is largely uniform and matches well with the set point.

In addition to establishing a relationship between the measured conversion and the quantitative chemical kinetics of the catalyst, it is also important to investigate the extent of any thermal gradients that may exist within the reactor. **Figure S5** displays the 3-dimensional temperature distribution in and around the *operando* ETEM reactor for four furnace thermocouple set points: of (a) 144 °C, (b) 202 °C, (c) 251 °C, and (d) 285 °C. Note that these correspond to conditions where the catalyst is active and producing $CO_2$. These temperatures were chosen to investigate as they correspond to those explored during the *operando* experiment.

Observe that the temperature distributions appear largely uniform in the hot zone of the reactor where the catalyst is located. These results are qualitatively similar to those reported previously (Mølgaard Mortensen et al., 2015; Vincent et al., 2020), and they are expected given the large thermal mass of the furnace holder. A quantitative comparison of the temperature difference between the furnace thermocouple set point and the average temperature at the TEM grid is presented below in **Table S2**. For the conditions explored here, the average temperature on the TEM grid surface differs from the set point by < 2 °C. The temperature at the grid is slightly less, as expected. Overall, these results indicate that the *operando* ETEM reactor can be treated as essentially isothermal and they suggest that the furnace thermocouple can be used as a reliable probe of the temperature surrounding the imaged catalyst.

**Table S2.** Quantitative investigation of the temperature distribution in the *operando* ETEM reactor. The furnace thermocouple set point is compared to the actual temperature at the TEM grid, which contains the catalyst that is imaged during an experiment. The maximum and minimum temperatures are given along with the average temperatures evaluated at the inner surface that coincides with the electron beam optic axis within the *operando* pellet.

| Furnace Set Point (°C) | Max Temperature at Inner Grid Surface (°C) | Min Temperature at Inner Grid Surface (°C) | Average Temperature at Inner Grid Surface (°C) | Difference between Set Point and Average Grid Temperature (°C) |
|---|---|---|---|---|
| 144 | 143.43 | 143.34 | 143.38 | 0.62 |
| 202 | 201.10 | 200.97 | 201.02 | 0.98 |
| 251 | 249.80 | 249.63 | 249.69 | 1.31 |
| 261 | 259.74 | 259.56 | 259.63 | 1.37 |
| 271 | 269.68 | 269.49 | 269.56 | 1.44 |
| 275 | 273.65 | 273.46 | 273.53 | 1.47 |
| 285 | 283.59 | 283.39 | 283.47 | 1.53 |
| 297 | 295.53 | 295.32 | 295.40 | 1.60 |

## 4. Time-averaged image methodology

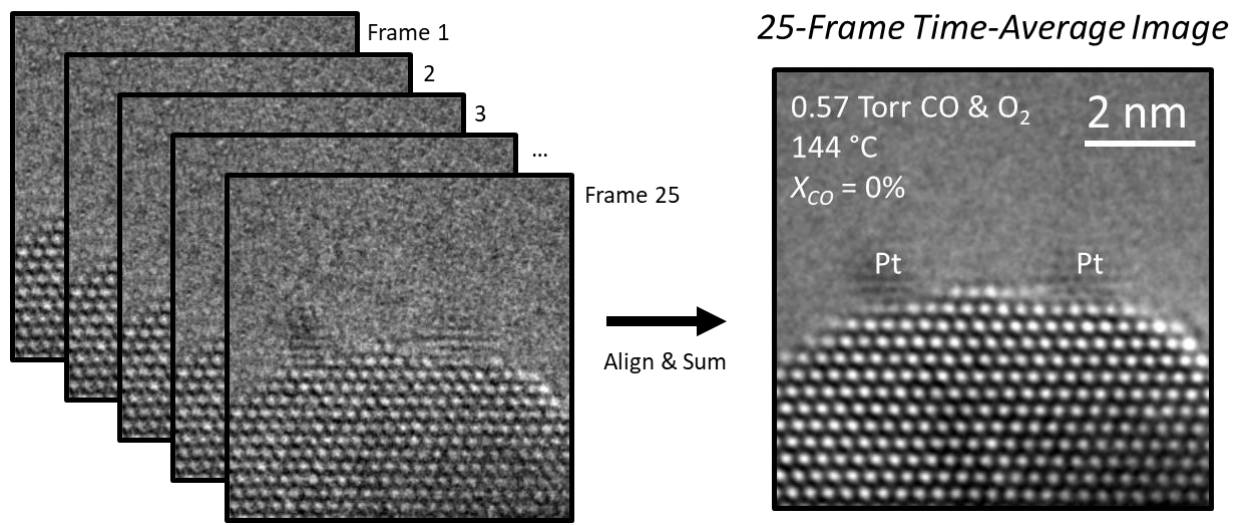

**Figure S6.** Schematic of time-averaged image methodology. The limited electron flux of about $10^3$ e$^-$ Å$^{-2}$ s$^{-1}$ used here results in a poor SNR in each individual 0.5 s exposure frame (left). To increase the SNR in the image without increasing the electron dose, multiple independent frames were selected, aligned, and summed together to produce a time-averaged image (right). Individual images were selected on the basis of bulk Ce column visibility to avoid artifacts from drift in the sample or electron optics. The time-average shown at right was constructed from 25 frames, which yields a 12.5 s time-average. The time-averaged image exhibits a marked increase in SNR available for local structural analysis.

## 5. Multislice TEM image simulations of contrast reversals in $CeO_2$-supported Pt nanoparticles undergoing rigid-body rotations

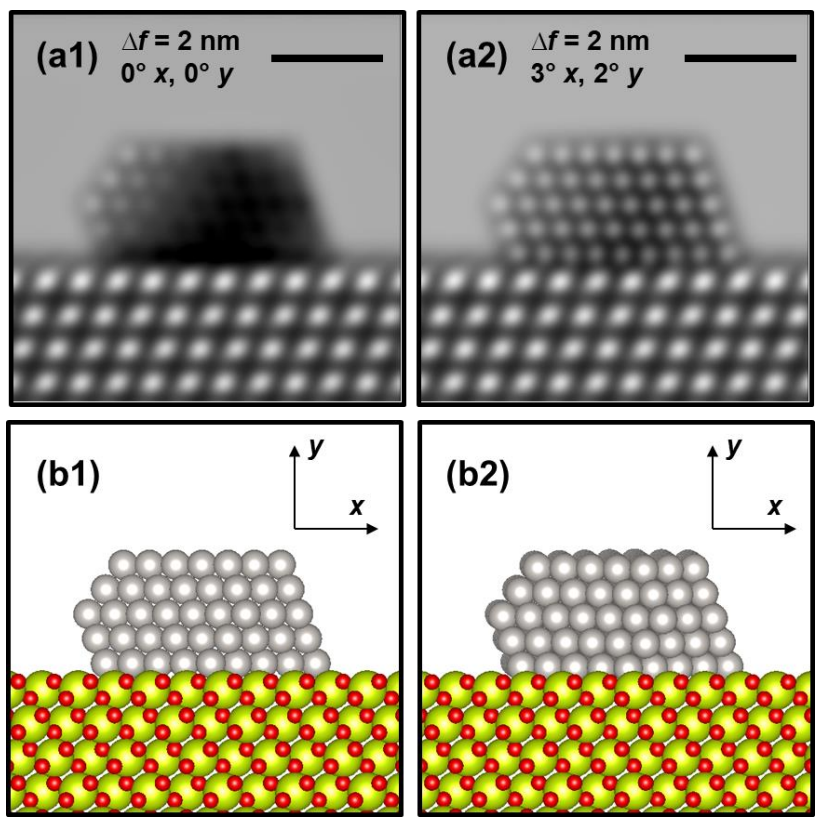

**Figure S7.** Multislice TEM image simulations of a Pt nanoparticle supported on a $CeO_2$ (111) surface demonstrating the appearance of mixed black and white Pt atomic column contrast and contrast reversals due to rigid body rotations. Figure **(a1)** presents a TEM image simulation of a Pt particle that is oriented along the incident electron beam without any tilt, calculated for an electron optical defocus of 2 nm. Notice the appearance of mixed black and white atomic column contrast in the Pt nanoparticle due to a contrast reversal occurring between the thinner sites on the edge of the particle and the thicker sites in the particle center. Figure **(a2)** presents another simulation at the same defocus but now for a Pt particle that has been tilted by three degrees about the horizontal $x$ axis and by two degrees about the vertical $y$ axis (see **(b1)** and **(b2)** for the atomic models as well as the orientation of the axes). In the case that the particle is tilted, the image simulated at a defocus of 2 nm no longer shows a mixture of black and white atomic column contrast but instead shows only white atomic column contrast, presumably due to a disruption in the electron beam channeling conditions. Simulations performed at other comparable tilts (i.e., of a few degrees) show similar behavior. Simulations were performed in the Dr. Probe software package (Barthel, 2018), using an accelerating voltage of 300 kV, a beam convergence angle of 0.2 mrad, a focal spread of 4 nm, a $3^{rd}$ order spherical aberration of -13 μm, a $5^{th}$ order spherical aberration coefficient of 5 mm, and a defocus of 2 nm. During the simulation, the $CeO_2$ support was tilted independently by 1.5° $x$ and 1.0° in $y$. The tilt information in the figure inset of **(a1)** and **(a2)** therefore denote the orientation of the Pt with respect to the incident electron beam. Scale bars in **(a1)** and **(a2)** correspond to 2 nm. An isotropic vibration envelope of 85 pm was applied during the image calculation. In the models, Pt is gray, Ce is yellow-green, and O is red.

## 6. Analysis of Bragg spots appearing in Fourier transform of Pt nanoparticles imaged under CO oxidation reaction conditions

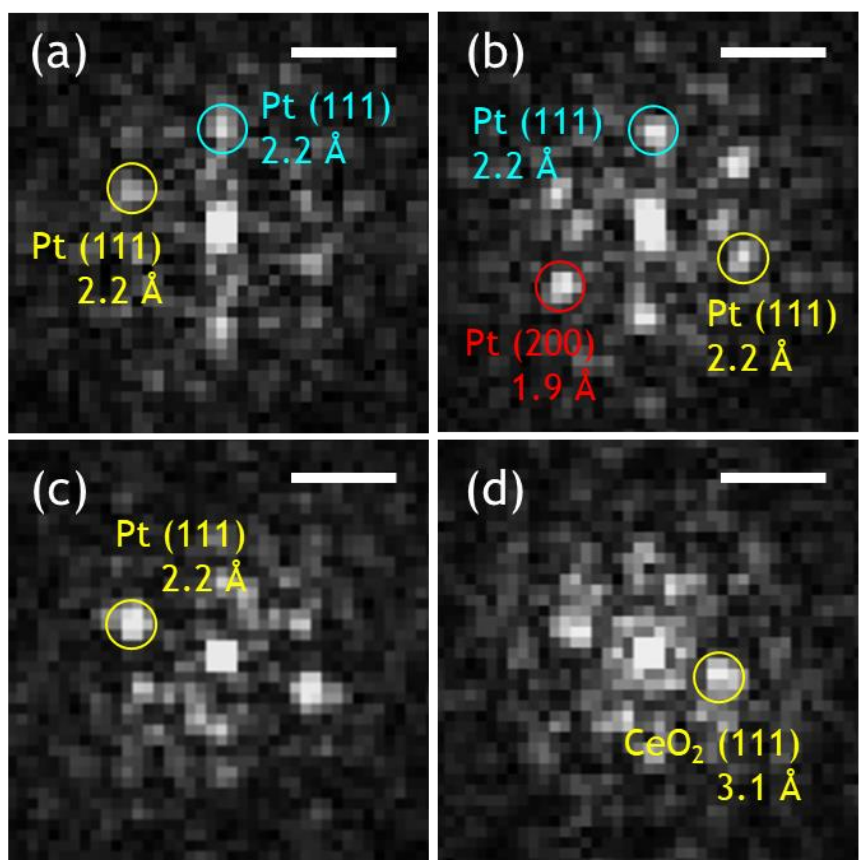

**Figure S8.** Indexing major Fourier transform (FT) spots shown in Figure 2(f1) - (f4) in the main text. Here, **(a)** corresponds to (f1), **(b)** to (f2), and so on. Major FT spots corresponding to the (111) and/or (200) lattice plane spacings of the Pt metal phase can be seen in **(a)**, **(b)**, and **(c)**, while in a spot corresponding to the $CeO_2$ (111) lattice plane spacing is discernable in **(d)**. Color coding is simply for clarity. The scale bar in each subfigure corresponds to 5.0 nm$^{-1}$.

## 7. Dynamic structural response of CeO$_2$-supported Pt nanoparticles at elevated temperatures

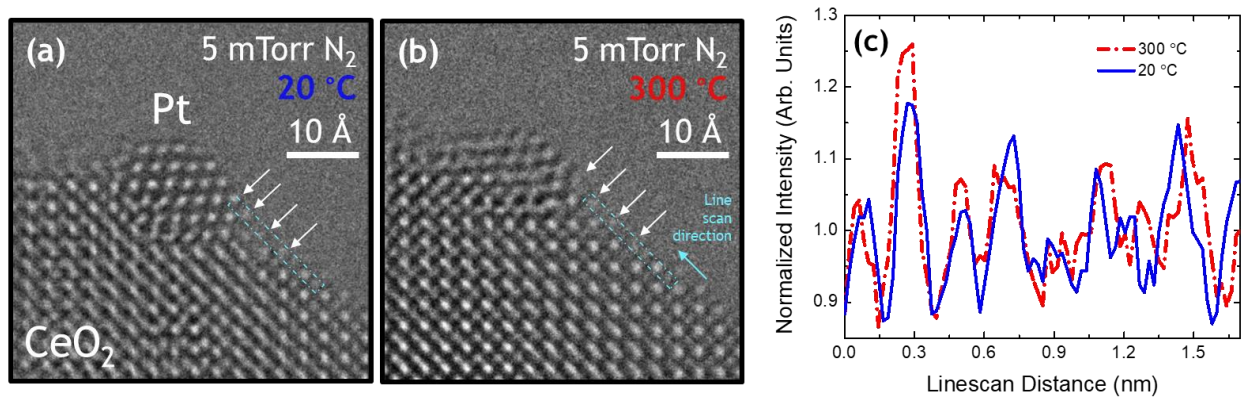

**Figure S9.** Half-second exposure *in situ* ETEM images of the same nanoparticle of a Pt/CeO$_2$ catalyst in 5 mTorr N$_2$ at **(a)** 20 °C and **(b)** 300 °C. Clearly resolved atomic columns bridging the Pt/CeO$_2$ three-phase boundary (TPB) are indicated with white arrows. In **(c)** a 1.2 Å wide integrated intensity line profile taken across the TPB from the region and along the direction indicated in the images shows that the atomic column contrast at 20 °C differs little from that observed when the catalyst is heated up to 300 °C. These results suggest that both the loss of Pt fringe visibility and the evolution of Ce column blurring observed during catalysis are not attributed strictly to thermal effects, but instead to dynamic chemical/structural processes resulting from catalytic turnover.

The analysis of the time-averaged *operando* TEM images shows that the turnover frequency for CO oxidation is correlated with structural dynamics taking place in the Pt nanoparticles and at/near the three-phase boundary (TPB). However, as the catalyst must be heated in order to be activated, it may be reasoned that the dynamics are driven by the elevated temperature of the catalyst and not the catalytic chemistry. We have therefore performed *in situ* TEM imaging experiments in inert gases at elevated temperatures, in order to investigate what may be called spectator fluxional behavior attributed only to the presence of applied heat and not to catalytic surface chemistry. **Figure S9** presents 0.5 s exposure *in situ* ETEM images of the same CeO$_2$-supported Pt nanoparticle in 5 mTorr of inert N$_2$ gas at (a) 20 °C and (b) 300 °C. Note that 300 °C is nearly the highest temperature explored in the *operando* ETEM experiment (297 °C).

Compared to the observations made *in situ* under reaction conditions (e.g., Figure 2 of the main text) or the case in which the catalyst is exposed to reactant gases and actively producing CO$_2$ (see e.g., Figure 4 of the main text), the catalyst exhibits a significantly different and more attenuated dynamic structural response as the temperature is elevated from 20 °C to 300 °C in an atmosphere of inert N$_2$. For instance, the Pt nanoparticles were observed to undergo fluxional dynamics that resulted in a loss of Pt lattice fringe visibility. As can be seen in **Figure S9**, at room temperature the Pt nanoparticle is well-oriented into a [110] zone axis. Clear atomic-column contrast is visible in the Pt as well as at the interfacial sites that bridge the right side of the Pt/CeO$_2$ perimeter (white arrows). The left side of the three-phase boundary yields contrast that is challenging to interpret due to thickness/tilt issues as well as the apparent overlap of the Pt and $CeO_2$ lattices. Nevertheless, the sharp visibility of the arrowed interfacial sites and Pt atomic columns indicates that the Pt nanoparticle, three-phase boundary, and nearby free $CeO_2$ surface exhibit general structural stability (fluxional dynamics, in comparison, would give rise to distinct motion artifacts, such as blurring or a localized loss of contrast). A 0.5 s exposure *in situ* ETEM image of the same Pt nanoparticle at an elevated temperature of 300 °C is shown in **Figure S9b**. At elevated temperature, the Pt nanoparticle has now tilted into an oblique orientation near the [100] zone axis. Regardless of the static restructuring that may have occurred (presumably to lower the overall system energy at this higher temperature), the primary observation of importance for the present study is the lack of blurring and/or localized motion artifacts associated with fluxional dynamics. In this case, multiple Pt lattice fringes remain sharply resolved in the nanoparticle and atomic columns can even be seen on the Pt surface. Images acquired at different times during the observation also show well-resolved Pt lattice fringes and sharp atomic columns (see, e.g., **Figure S10**), suggesting that the loss of Pt lattice fringe visibility observed under *operando* conditions is attributed to structural dynamics driven by catalytic surface chemistry. A close inspection of the images reveals that the particle occasionally undergoes rigid-body rotations of a few degrees (compare, e.g., **Figure S9b** with **Figure S10b**). Notably, such behavior was also seen to occur at room temperature (compare, e.g., **Figure S9a** with **Figure S10a**). This type of slow rotational behavior is completely different than the rapid dynamic structural fluctuations that we have observed during the experiments reported in the main text (see, e.g., Figure 2 or Figure 4) or during *in situ* TEM imaging experiments that we have performed of $CeO_2$-supported Pt nanoparticles exposed to CO (Crozier et al., 2019), CO and $O_2$ (Vincent & Crozier, 2020), and CO and $H_2O$ (Li et al., 2021).

A second dynamic structural behavior observed during the *operando* TEM experiment was the fluxionality of Ce atomic columns located at the three-phase boundary (TPB) and along the nearby free $CeO_2$ support surface. Here, we observe in inert $N_2$ that the atomic columns located at/near the three-phase boundary show similar contrast at 20 °C and 300 °C. For example, in **Figures 9a/b** and **S10a/b**, white arrows have been used to indicate a series of atomic columns that bridge the free $CeO_2$ support surface across the TPB to the Pt NP surface. Close inspection of the images and an examination of the intensity profile across this region reveal that the contrast and the width of the atomic columns are alike at low and elevated temperature. **Figure 9c** presents integrated intensity line profiles taken from a 1.2 Å wide window across the TPB from the region indicated by the dashed box in the corresponding images. The intensity line profiles have been normalized to the average vacuum intensity; the profile from 20 °C is plotted as a solid blue line, while the one from 300 °C is plotted as a dashed red line. A direct comparison of the magnitude and the shape of the profiles demonstrates that the contrast of the atomic columns at/near the Pt/$CeO_2$ TPB remains relatively unchanged as the catalyst is heated from 20 °C to 300 °C in $N_2$. These results suggest that the blurring of atomic sites at/near the Pt/$CeO_2$ TPB observed during the operando TEM experiment arises from structural dynamics which do not occur in the absence of reactants.

Finally, we investigate the lattice plane separation distance of the top layer on the $CeO_2$ support near the $Pt/CeO_2$ interface. The surface lattice plane separation was measured from intensity line profiles taken from the bulk of the $CeO_2$ toward the surface; the region over which the profiles were taken at 20 °C and 300 °C are indicated in **Figure S11a/b**, and the intensity profiles are themselves given in **Figure S11c**. Measurements of the surface and the sub-surface lattice plane separation distance yield 1.98 ± 0.1 Å at both temperatures. Here we note that these measurements match those obtained from the bulk of the nanoparticle and that they agree with the accepted $CeO_2$ (110) Miller plane spacing of 1.91 Å. Thus, the outward $CeO_2$ surface relaxation observed under *operando* conditions also appears to be a consequence of structural dynamics that are driven by catalytic chemistry and not heating to 300 °C alone.

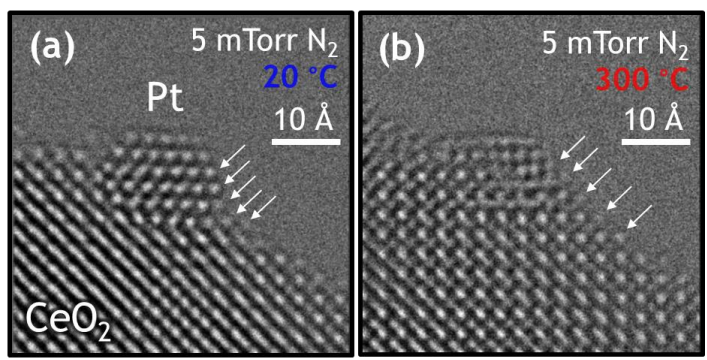

**Figure S10.** Additional half-second exposure *in situ* ETEM images of the same nanoparticle of a $Pt/CeO_2$ catalyst in 5 mTorr $N_2$ at **(a)** 20 °C and **(b)** 300 °C. Clearly-resolved atomic columns bridging the $Pt/CeO_2$ TPB are indicated with white arrows.

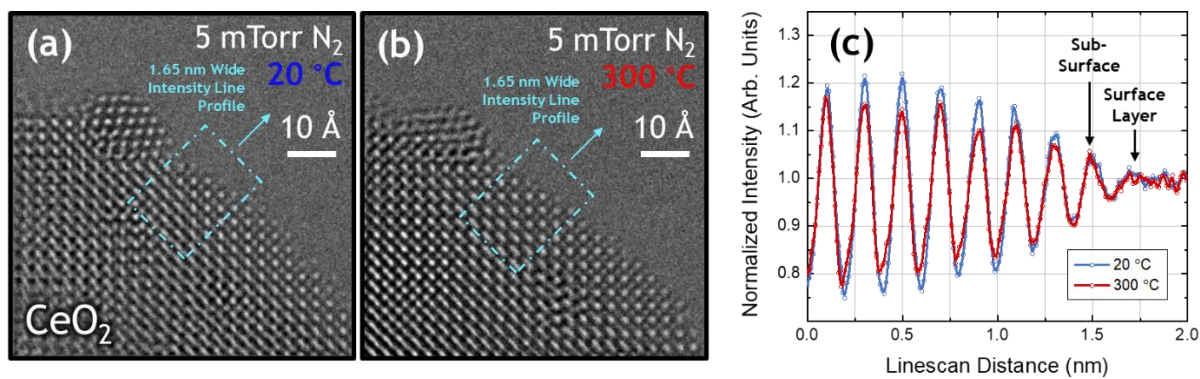

**Figure S11.** Half-second exposure *in situ* ETEM images of the same nanoparticle of a $Pt/CeO_2$ catalyst in 5 mTorr $N_2$ at **(a)** 20 °C and **(b)** 300 °C, which were presented in Figure S8 and which are shown here again to indicate the regions over which integrated intensity line profiles were taken to measure the surface lattice plane separation distance. The regions are indicated with light blue dashed boxes; the profiles were taken in the arrowed direction which lies perpendicular to the $CeO_2$ (110) surface. In **(c)** the intensity line profiles at 20 °C (blue line) and 300 °C (red line) are plotted overtop one another. The line profiles were normalized to the average vacuum intensity; original data points are plotted as circles along with an interpolated spline function for clarity. The surface layer and the sub-surface lattice plane are indicated in the graph.

## 8. Additional *in situ* ETEM image time-series of CeO$_2$-supported Pt nanoparticles under CO oxidation reaction conditions

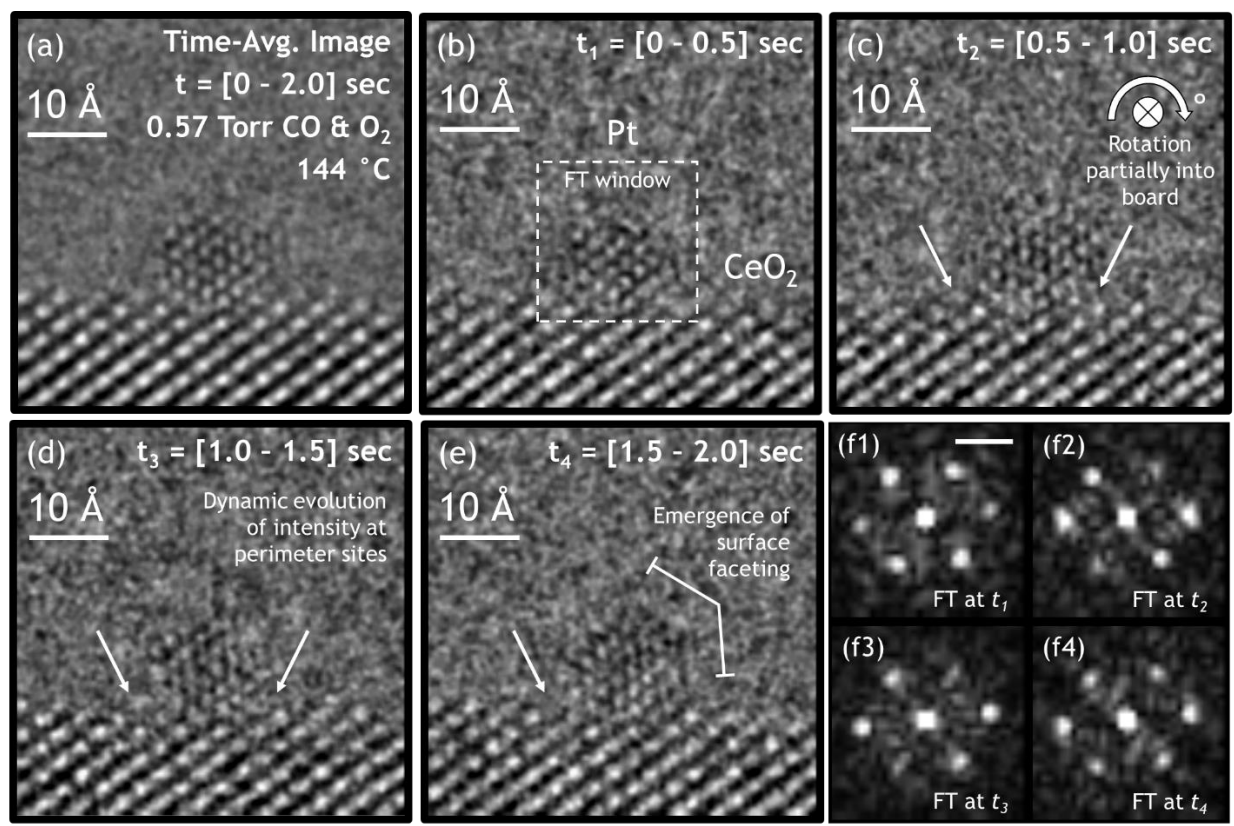

**Figure S12.** *In situ* ETEM image time-series of CeO$_2$-supported Pt NP at 144 °C in 0.57 Torr of CO and O$_2$. Part **(a)** shows the time-averaged image of the catalyst, obtained by summing together the individual 0.5 second exposure frames over the entire over [0 – 2.0] second acquisition period. Parts **(b)** – **(e)** show the atomic-scale structural dynamics that evolve over 0.5 second intervals from $t = 0$ seconds to $t = 2.0$ seconds. Parts **(f1)** – **(f4)** display the FT taken at each time interval from the windowed region around the Pt NP, as denoted in (b). The scale bar in (f1) is 5.0 nm$^{-1}$. Images have been processed with a bandpass filter for clarity. FTs were produced from unfiltered, windowed images that were processed with a Hanning function to remove edge artifacts caused by windowing; the modulus of the FT is shown.

## 9. Analysis of intensity line profiles to quantify Ce atomic column blurring and CeO$_2$ (111) outward surface relaxation

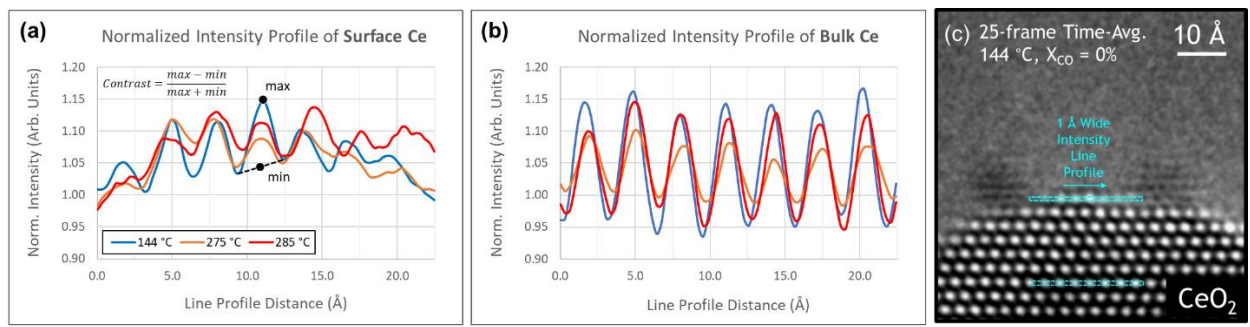

**Figure S13.** Normalized intensity line profiles taken from **(a)** the free surface Ce columns between the Pt nanoparticles and from **(b)** a row of Ce cation columns several layers into the bulk of the nanoparticle. All of the intensity line profiles have been normalized to the average vacuum intensity at each condition. Intensity profiles are given for 144 °C (blue), 275 °C (orange), and 285 °C (red). The line profiles were generated over an integration window 92.1 pm wide (i.e., 3 pixels). The profile windows and direction are indicated in **(c)**. In (a) the equation for calculating the contrast of a column is given, and the process for determining the maximum and minimum values is demonstrated for the middle surface Ce column at 144 °C. In (b) the fact that the CeO$_2$ particle has tilted at 275 °C is evident by the reduction in contrast of the orange curve. When the catalyst was heated to 285 °C, the crystal tilted back, seen by the recovered contrast in the red curve.

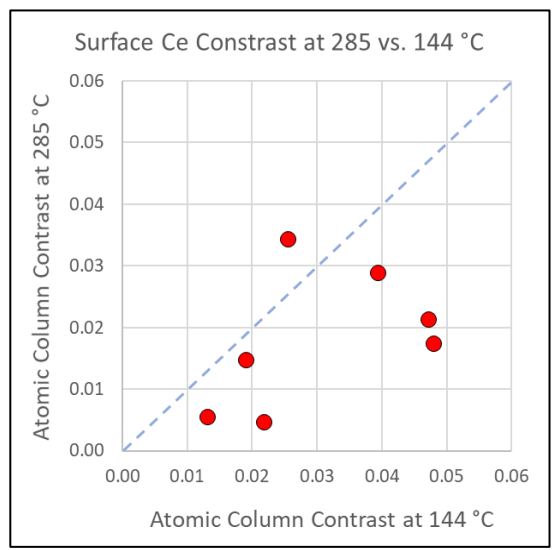

**Figure S14.** The contrasts of the surface Ce columns at 24% CO conversion are plotted against the contrasts of the same column at 0% CO conversion. A straight dashed line is provided for reference, which shows that most Ce columns become more blurred with increasing conversion, as most points lie below the straight dashed line.

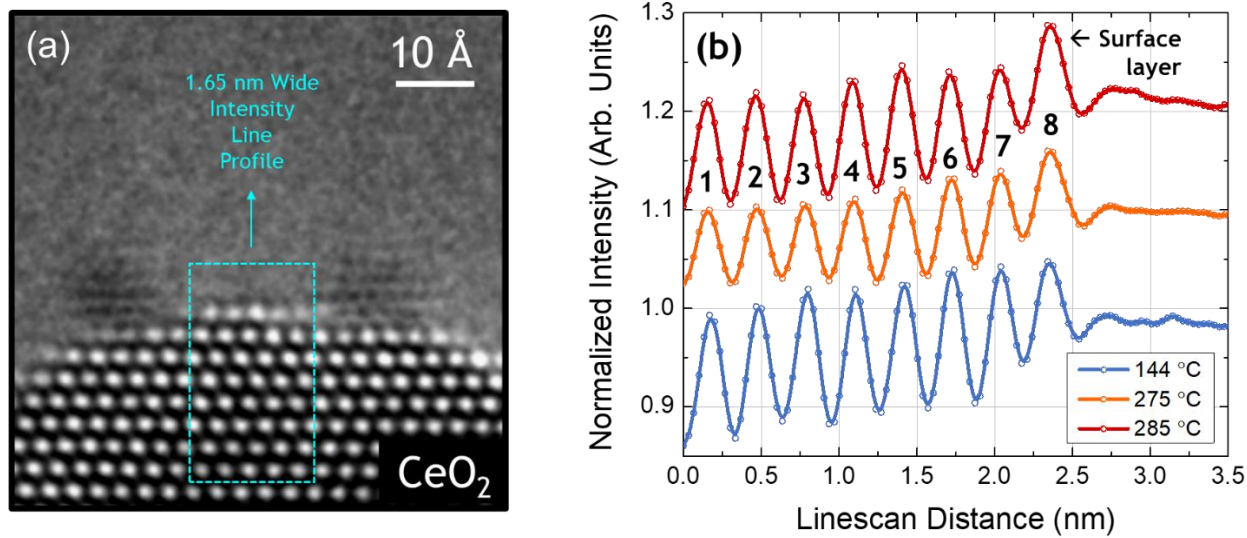

**Figure S15.** Normalized intensity line profiles taken from interior of nanoparticle toward catalyst surface, revealing the outward $CeO_2$ (111) surface relaxation with increasing catalytic turnover. In **(a)** the region and direction of the intensity line scans are indicated for an example *operando* TEM time-averaged image acquired at 144 °C. The same region was used for all three conditions. In **(b)** the normalized intensity line profiles are plotted for 144 °C (blue), 275 °C (orange), and 285 °C (red). Original data points are plotted along with an interpolated spline which was used to measure the distances between the peaks in the profile. Numerical labels are given to the peaks to identify the measured separation distances that are provided in Table S3.

**Table S3.** Measurement of peak separation distances from intensity line profiles presented in Figure S15. The peaks being measured are referenced with the numerical labels given in Figure S15b (top). The outward relaxation of the $CeO_2$ (111) surface layer is presented in the last row of the table. The standard deviation of the measurements taken of all layers except the bulk (second to last row of the table) is used as an error bar on the measurement at each condition. Note that the averages of the sub-surface and bulk layer separation distances (third to last row of the table) are all close to the accepted $CeO_2$ (111) Miller plane spacing of 312 pm.

| Peaks Being Measured | Separation Distance at 144 °C (pm) | Separation Distance at 275 °C (pm) | Separation Distance at 285 °C (pm) |
|---|---|---|---|
| 1 – 2 | 305 | 313 | 307 |
| 2 – 3 | 313 | 307 | 306 |
| 3 – 4 | 308 | 313 | 310 |
| 4 – 5 | 311 | 309 | 315 |
| 5 – 6 | 303 | 314 | 307 |
| 6 – 7 (sub-surface) | 312 | 312 | 319 |
| Average of all layers except the surface | 309 | 311 | 311 |
| Std. dev. of all layers except the surface | 4 | 3 | 5 |
| **7 – 8 (surface layer)** | **306** | **315** | **320** |

# References

BARTHEL, J. (2018). Dr. Probe: A software for high-resolution STEM image simulation. *Ultramicroscopy* **193**, 1–11.

CROZIER, P. A., LAWRENCE, E. L., VINCENT, J. L. & LEVIN, B. D. A. (2019). Dynamic Restructuring during Processing: Approaches to Higher Temporal Resolution. *Microscopy and Microanalysis* **25**, 1464–1465.

ERTL, G., KNZINGER, H. & WEITKAMP, J. (eds.) (1997). *Handbook of Heterogeneous Catalysis*. Weinheim, Germany: Wiley-VCH Verlag GmbH http://doi.wiley.com/10.1002/9783527619474.fmatter (Accessed August 31, 2020).

LI, Y., KOTTWITZ, M., VINCENT, J. L., ENRIGHT, M. J., LIU, Z., ZHANG, L., HUANG, J., SENANAYAKE, S. D., YANG, W. D., CROZIER, P. A., NUZZO, R. G. & FRENKEL, A. I. (2021). Dynamic structure of active sites in ceria-supported Pt catalysts for the water gas shift reaction. *Nature Communications* **12**, 914.

MAI, H. X., SUN, L. D., ZHANG, Y. W., SI, R., FENG, W., ZHANG, H. P., LIU, H. C. & YAN, C. H. (2005). Shape-selective synthesis and oxygen storage behavior of ceria nanopolyhedra, nanorods, and nanocubes. *Journal of Physical Chemistry B* **109**, 24380–24385.

MØLGAARD MORTENSEN, P., WILLUM HANSEN, T., BIRKEDAL WAGNER, J. & DEGN JENSEN, A. (2015). Modeling of temperature profiles in an environmental transmission electron microscope using computational fluid dynamics. *Ultramicroscopy* **152**, 1–9.

VINCENT, J. & CROZIER, P. (2020). Atomic-resolution *Operando* and Time-resolved *In Situ* TEM Imaging of Oxygen Transfer Reactions Catalyzed by CeO2-supported Pt Nanoparticles. *Microscopy and Microanalysis* **26**, 1694–1695.

VINCENT, J. L., VANCE, J. W., LANGDON, J. T., MILLER, B. K. & CROZIER, P. A. (2020). Chemical Kinetics for Operando Electron Microscopy of Catalysts: 3D Modeling of Gas and Temperature Distributions During Catalytic Reactions. *Ultramicroscopy* 113080.